\documentclass[12pt,preprint]{aastex}
\usepackage{graphicx}
\usepackage{natbib}
\usepackage{changepage} 
\def\swift{{\em Swift\/}}
\def\sax{{\em BeppoSAX\/}}
\def\gro{{\em Compton Gamma Ray Observatory\/}}   
\def\fermi{{\em Fermi\/}}
\def\batse{{\em BATSE\/}}

\makeatletter
\def\@cite#1#2{(#1\if@tempswa , #2\fi)}
\def\preprint{preprint}   \newif\ifPreprintMode
\ifx\preprint\revtex@genre\PreprintModetrue\else\PreprintModefalse\fi
\makeatother

\begin{document}

\title{Broad band time-resolved $E_{p,i}$--$L_{\rm iso}$ correlation in GRBs}
\author{F.~Frontera\altaffilmark{1,2},
L.~Amati\altaffilmark{2},
C.~Guidorzi\altaffilmark{1},
R.~Landi\altaffilmark{2},
J.~in't Zand\altaffilmark{3}
}
\altaffiltext{1}{Universit\`a di Ferrara, Dipartimento di Fisica, Via Saragat 1, 
44100 Ferrara, Italy; email: frontera@fe.infn.it}
\altaffiltext{2}{INAF,Istituto di Astrofisica Spaziale e Fisica Cosmica,
Bologna, Via Gobetti 101, 40129 Bologna, Italy}
\altaffiltext{3}{Space Research Organization in the Netherlands,
 Sorbonnelaan 2, 3584 CA Utrecht, The Netherlands}

\begin{abstract}
We report results of a systematic study of the broad band (2--2000~keV) time resolved 
prompt emission spectra of a sample of gamma-ray bursts (GRBs) detected with both Wide 
Field Cameras on board the \sax\ satellite and the \batse\ experiment  on board CGRO. 
In this first paper, we study the time-resolved dependence of the 
intrinsic peak energy $E_{p,i}$ of the $E F(E)$ spectrum on the corresponding isotropic 
bolometric luminosity $L_{\rm iso}$. The $E_{p,i}$--$L_{\rm iso}$ relation or the equivalent relation 
between $E_{p,i}$ and the bolometric released energy $E_{iso}$, derived using the time 
averaged spectra of long GRBs with known redshift, is well 
established, but its physical origin is still a subject of discussion. In addition, 
some authors maintain that these relations are the result of instrumental selection effects.
We find that not only a relation  between the measured peak energy $E_p$ and the 
corresponding energy flux, but also a strong  $E_{p,i}$ versus
$L_{\rm iso}$ correlation are 
found within each burst and merging together the time resolved  data points from 
different GRBs. 
 We do not expect significant instrumental selection effects that can affect the 
obtained results,
apart from the fact that the GRBs in our sample are sufficiently bright to perform a 
time-resolved spectroscopy and that they have known redshift.
If the fundamental  physical process that gives rise to the GRB phenomenon does not depend 
on its brightness, we
conclude that the found $E_{p,i}$ versus $L_{\rm iso}$ correlation within each GRB is intrinsic to the emission process,  
and that the correlations discovered by Amati et al. and 
Yonetoku et al. are likely not the result of selection effects. 
We also discuss the properties of the correlations found.
\end{abstract}

\keywords{gamma ray bursts: general --- gamma rays: stars}

\maketitle

\section{Introduction}
\label{intro}

In spite of the major advances in knowledge of gamma--ray bursts (GRBs) afterglow properties, 
mainly made with \swift, the GRB phenomenon is still poorly understood \citep{Lyutikov09}. 
It is recognized that the study of the prompt emission is of crucial importance, as it is more 
directly connected with the original explosion. 
One of the issues still open is the radiation emission mechanism(s) at work. 
Most of the GRB properties derived thus far come from the
time-averaged spectra.
The function that has been found to better describe them from 15 keV up to 10 MeV  
is a smoothly broken power-law proposed by \citet[{\em Band function}, BF]{Band93}.
%
On the basis of the spectral data obtained with the {\em Burst and Transient 
Source Experiment} (\batse), aboard the \gro\ satellite
({\em CGRO}) and, for example,  
with the \sax\ GRBM data \citep[e.g.,][]{Guidorzi11}, for long GRBs ($>$2 s), the mean value 
of the low-energy photon index $\alpha$ of the BF is about $-1$, while that of the 
high-energy photon index $\beta$ is about $-2.3$ \citep{Kaneko06}. 
As a consequence of this result, the received power per unit logarithmic 
energy interval  $EF(E)$ shows a peak value that, in the \batse\ era, seemed to show a sharp 
Gaussian distribution around 200 keV. With the discovery of the X--ray flashes with 
\sax, later also found with {\em HETE}-2, \swift, and, now, with the \fermi\ Gamma-ray Burst 
Monitor, this distribution results in being much 
flatter \citep[e.g.,][]{Kippen03,Sakamoto05}.  In the cases in which $\beta$ cannot 
be constrained, a power-law model with a high energy exponential cutoff ({CPL) 
gives a good fit to the data, and, in some cases, even a simple power-law 
can describe the GRB time averaged spectra up to several MeV photon 
energies \citep{Kaneko06}.

Several radiative emission models have been worked out for the interpretation of the GRB
spectra. Given their non-thermal shape, the first model proposed was 
synchrotron emission by non thermal electrons
in strong magnetic fields \citep{Rees94,Katz94,Tavani96}. Indeed, the time-averaged 
spectra of many GRBs are consistent with an optically thin synchrotron shock model 
\citep[e.g.,][]{Tavani96,Amati01}. However, there are a significant number of GRBs for which
this model does not work. Indeed, while for an optically thin synchrotron
spectrum, the expected power-law index of the $E F(E)$ spectrum below the
peak energy $E_p$ cannot be steeper than 4/3 (ideal case of an instantaneous
spectrum in which electron cooling is not taken into account), in 
many cases \citep[e.g.,][]{Preece98,Preece00}
the measured spectra, even those time resolved \citep{Crider97,Frontera00a}, 
are inconsistent with these expectations. 

To overcome these difficulties, either modifications of the above synchrotron scenario
\citep[e.g.,][]{Lloyd00a} or other radiative models 
\citep[e.g.,][]{Liang97,Blinnikov99,Lazzati00,Meszaros00,Stern04,Peer06,Peer07,Lazzati09,Peer11}
have been suggested. Each of these models interprets some of  
the prompt emission features, but fails to interpret others.  

One of the GRB spectral properties that the emission models should  interpret
is the correlation between the intrinsic (redshift corrected) peak energy $E_{p,i}$ of 
the $E F(E)$ function and either the GRB released energy $E_{{\rm iso}}$ \citep{Amati02} or the 
peak bolometric luminosity $L_{p,{\rm iso}}$ \citep{Yonetoku04}. Both correlations 
(the Amati relation and the Yonetoku relation) have been derived from the time-integrated 
spectra assuming  isotropic emission. The Yonetoku relation followed the Amati's, from which 
the so-called Ghirlanda relation
\citep{Ghirlanda04a} was also derived by replacing the released energy $E_{\rm iso}$ with that 
($E_{\gamma}$) corrected for the beaming factor ($E_{\gamma} = (1-\cos{\theta}) E_{\rm iso}$).
The latter is model dependent, being derived assuming a jet like structure of the 
fireball, a constant
efficiency of the fireball in converting kinetic energy in the ejecta into gamma rays,
a mass density  distribution of the circumburst medium, and, mainly, that the break time 
observed in the late afterglow light curve occurs when reciprocal of  
the bulk Lorentz factor of the jet, $1/\Gamma$, becomes of the order of the jet opening
angle $\theta_{\rm jet}$. The latter assumption requires that the break time is achromatic,
property not observed in many \swift\ GRBs \citep[e.g.,][]{Campana07,Melandri08}. 
In any case, these
relations are all equivalent as far as the physics: a relation between
the photon energy in which most of the energy is released and the electromagnetic 
radiation emitted by a GRB, which can be expressed equivalently in terms of 
released energy or peak (or average) GRB luminosity.    

The Amati relation \citep[$E_{p,i} = K E_{\rm iso}^m$, with $K= 98\pm 7$ and $m = 0.54\pm0.03$,
where $E_{p,i}$ is measured in keV and $E_{\rm iso}$ is given in units of $10^{52}$~erg]{Amati08},   
is satisfied, within an extra-Poissonian scatter of $\log{E_{p,i}}$ normally distributed around
the best-fit power-law with $\sigma\sim 0.2$~dex, by all long GRBs (more than 100) 
with known redshift $z$ discovered thus far, except the nearest and least energetic 
GRB ($z= 0.0085$) ever observed (GRB\,980425) and, maybe, GRB\,031203 \citep[e.g.][]{Amati07}, 
but not by short GRBs. Also the Yonetoku relation 
is satisfied for GRBs with known redshift, except GRB\,980425 
(e.g., \citet{Ghirlanda05b,Nava12}), given the tight correlation found 
between $L_{p,{\rm iso}}$ and $E_{\rm iso}$ \citep{Ghirlanda05b}.

In spite of this, the Amati relation has been questioned by various 
authors \citep{Band05,Butler07b,Butler09,Shahmoradi09,Collazzi11}, maintaining that it is likely 
the result of selection effects, even if these effects, when investigated by other authors 
\citep{Ghirlanda05a,Ghirlanda08,Nava08,Amati09,Krimm09,Nava11a} are found
to be marginal.

One of the major criticisms of the Amati relation is that its normalization  depends
on instrument sensitivity \citep{Butler07b}. However, this conclusion by 
\citet{Butler07b} for \swift\ GRBs was not based on measured spectral peak energies 
$E_{p,i}$, but on $E_{p,i}$ values inferred, under some assumptions, from a Bayesian method.  
An investigation, performed by \citet{Amati09}  by deriving the $E_{p,i}$ dependence on 
$E_{\rm iso}$ for different sets of GRBs, each obtained with a different instrument, and by using
the measured values of $E_{p,i}$ reported in the \swift\ BAT 
official catalog by \citet{Sakamoto08}, has not confirmed the inferences by \citet{Butler07b}. 

Other works that question the reliability of the $E_{p,i}$ -- $E_{\rm iso}$ make use of 
the ($E_p$, Fluence) observer plane, positioning in this plane the data points obtained 
from the spectral analysis of GRBs detected with different satellite instruments, using 
the method first
proposed by \citet{Nakar05}. While some authors \citep[e.g.][]{Band05,Goldstein10} find 
that most of the long \batse\ GRBs do not satisfy the relation, other authors 
\citep[e.g.,][]{Ghirlanda05a,Nava08,Nava11a} find  that most of them do. 
This discrepancy is due to various reasons, such as the condition assumed to satisfy 
or not satisfy the Amati relation  and the systematic errors in the determination of
fluence and $E_p$ 
\citep[e.g.][]{Collazzi11}. 
However some results are well established and are shared by most authors who have 
performed the $E_{p,i}$ -- $E_{\rm iso}$ test. GRBs with known redshift are not a 
special sub-population, they are evenly distributed along the weaker correlation between 
$E_p$ and fluence \citep{Ghirlanda08,Collazzi12}. No evidence of evolution of the
$E_{p,i}$ -- $E_{\rm iso}$ correlation with the redshift is found \citep{Ghirlanda08,Nava11a}. 
In the observer frame, after all the results obtained with \swift\ and \fermi, 
no bursts with large fluence and low/intermediate $E_p$ have been found. They do 
not exist or should be very rare: Nothing prevents their detection. Instead,  bursts with 
intermediate/high $E_p$ and small fluence could be affected by instrumental selection 
effects, such as the minimum flux to trigger a GRB, the minimum fluence to fit its spectrum 
and constrain its $E_p$, and truncation effects related to the instrument passband \citep{Lloyd00b}. 
The case of \swift\ BAT is an example of instrument affected by truncation biases: 
In spite of its higher fluence sensitivity than \batse, $E_p$ can be accurately 
determined only if it has a value within or very close to its energy passband (15--150 keV). 
However, we find some results about the validity of  the $E_{p,i}$ -- $E_{\rm iso}$ 
relation untenable, such as the conclusion 
%
%
by \citet{Collazzi12}, who find 
that a significant fraction of the same GRBs with known redshift which have 
been used to derive the Amati relation do not satisfy their test. Clearly, their 
test condition (their so-called "Amati limit") in the (Fluence, $E_p$) plane 
is too restrictive. The same authors state that, while the Amati relation is the result of
selection effects, the Ghirlanda relation is valid. Also, this statement is problematic, given
that this relation is based on the same $E_p$ and Fluence measurements, being the only correction 
performed with, as discussed above, the replacement of 
$E_{\rm iso}$ with $E_{\gamma}$. In addition, the "Ghirlanda limit" is derived using rather loose assumptions about the beaming angle. 

%
The opposite conclusion was recently reached by 
\citet{Yonetoku10} via analyzing in 
detail all possible data truncation and detector sensitivity effects, and by \citet{Nava12},  
who analyzed a complete sample, for redshift determination, of  bright \swift\ GRBs (1~s peak photon flux 
$P \ge 2.6$~photons~s$^{-1}$~cm$^{-2}$ in the 15--150 keV  BAT band). 
%

From this long standing debate, it is apparent the need to explore other approaches for testing 
the origin of the spectrum--energy correlations, i.e., whether they are the result of  
instrumental selection effects or are related to the fundamental physics of the 
GRB phenomenon. Given the significant evolution 
of the GRB spectra, studies of  time-resolved spectra 
are crucial not only to test the $E_{p,i}$ versus 
$L_{\rm iso}$ relation, but also  to delve deeper 
into the issue of the radiative mechanisms at work during the prompt emission.

Motivated by both these needs, we performed a systematic study of the broad band (2--2000~keV) 
time--resolved prompt emission spectra of a sample of GRBs detected 
with both Wide Field Cameras (WFCs) aboard the \sax\ satellite and the \batse\ 
experiment aboard the {\em CGRO}. 
In this paper we will concentrate on the test of the $E_{p,i}$ versus $L_{\rm iso}$ relation. 
The WFCs were among the few instruments  that detected GRB prompt emission down to 2 keV. 
Thus we can obtain time-resolved spectra in  an energy band still not well explored: the 
2--2000 keV band.
A paper devoted to testing physical emission models of GRBs using the same time-resolved spectra
will be the subject of a forthcoming paper (Frontera et al. 2012, in preparation).

\section{The GRB sample and spectral analysis}
\label{analysis}
There were nine GRBs simultaneously detected with WFCs and \batse: 970111, 971206, 971214, 
980329, 980519, 990123, 990510, 990907, 991030. 
We performed an analysis of the prompt emission spectra of all of them.
However, only four of them (970111, 980329, 990123, and 990510) were sufficiently 
bright to allow a fine time resolved spectroscopy (see Table~\ref{t:grb-intervals}). 

The instrumentation that detected these bursts is widely described in the literature.  
For the \batse\ experiment, see, e.g., \citet{Fishman94}, while, for WFC,  see \citet{Jager97}. 
The \batse\  spectra were taken from the Large Area Detectors (LADs), whose
typical passband is 25--2000 keV. The LADs provide various types of data products.
The data used for this analysis are the high energy resolution burst data, that
provide 128 energy channels with a minimum integration time of 64 ms. For details of the
\batse\ spectral data products and detector response matrix see \citet[and references therein]{Kaneko06}.

The WFCs consisted of two coded aperture cameras, each with a field
of view of 40$^\circ \times 40^\circ$ (full width at zero response) and
an angular resolution of 5 arcmin. They operated in normal mode with 31 energy 
channels in 2--28~keV and 0.5~ms time resolution.

The background-subtracted light curves of the four strongest GRBs in our sample, 
detected with both \batse\ and WFCs, are 
shown in Figure~\ref{f:lc}. For the \batse\ data, the background level was estimated using 
the count rates immediately before and after the GRBs. Given that the background is
variable during the GRB, it was  estimated by means of a parabolic interpolation, channel
by channel, between the background measured before the event and that measured after the event.
For WFC spectra, the background level was estimated 
using an equivalent section of the detector area not illuminated by the
burst or by other known X-ray sources. We also checked the consistency
of this background level with that obtained by using the data before and
after the burst. 

We subdivided the time profile of each GRB into a number of time slices (see Figure~\ref{f:lc}), 
taking into account the GRB profile as observed 
with 1 s time resolution (visible pulses, their rise, peak, and decay) and 
the count statistics. We performed 
further spectral analysis in all of the time slices in which it was possible to constrain 
the peak energy $E_p$. With reference to Figure~\ref{f:lc}, in the case of GRB\,970111 we excluded the first two (nos. 1 and  2) 
time slices for their low count statistics. For the same reason, we excluded from the 
analysis the first (no. 1) and the last (no. 8) of
GRB\,980329, the first (no. 1) of 
GRB\,990123, and intervals 3, 4, 5, and 6  of GRB\,990510. For this GRB, we also 
excluded intervals 13, 14 and 15, given that only an upper limit
could be obtained for $E_p$. In the case of GRB\,990123, we also excluded from the 
analysis the time intervals from 21 to 26, given that this part of the event was 
observed by WFC through the Earth's atmosphere. The number of time intervals in which 
we subdivided the time profile of each event, the number of selected time slices in 
which it was possible to estimate and constrain $E_p$, the GRB fluence, and its 
redshift when known, are given 
in Table~\ref{t:grb-intervals}. 
 
For each WFC$+$\batse\ time resolved spectrum we used as 
an input model the BF:
$$N(E) =  A \left(\frac{E}{100\,{\rm keV}}\right)^\alpha
        \exp{\left({-E/E_{0}}\right)}$$ \\
if $$(\alpha - \beta)\cdot E_{0} \hbox{  }\ge\hbox{  } E$$ \\
and $$N(E) = A \left[\frac{(\alpha - \beta) E_{0}}{100\,{\rm keV}}
\right]^{\alpha - \beta} \exp{(\beta - \alpha)}
\cdot \left(\frac{E}{100\,{\rm keV}}\right)^\beta$$ \\
if $$(\beta - \alpha) \cdot E_{0}\hbox{  }\le \hbox{  } E$$

where $\alpha$ and $\beta$ are the power law
low-energy (below E$_0$) and high-energy (above $E_0$) photon indices,
respectively, and $A$ is the normalization parameter. 

In the fit, along with  $A$, $\alpha$, and $\beta$, we adopted as a free parameter, 
instead of $E_0$, the photon peak energy $E_p = E_0 (2+\alpha)$. In addition, a normalization 
factor 
between \batse\ and WFC data was included in the fit and left to vary in the range 0.8--1.2, 
to take into account a possible intercalibration error. Actually, we found that, 
for all analyzed GRB spectra, this parameter was consistent with 1.  
The systematic error used by the
\batse\ team to take into account the uncertainty in the background subtraction and 
the uncertainty in the instrument response function (see, e.g., \citet{Kaneko06} for details)
was injected in the fit.
%

The input model was assumed to be photoelectrically absorbed (WABS model in XSPEC).  
Given that the absorption column density $N_{\rm H}$ could not be constrained, 
a Galactic absorption along the GRB direction \citep{Dickey90} was assumed. 
To deconvolve the count rate spectra we adopted the {\sc xspec} ({\em v. 12.5}) 
software package \citep{Arnaud96}.  
If not explicitly stated, the quoted uncertainties are single parameter 
errors at the 90\% confidence level.

\section{Results}

The fit results of the BF to the joint WFC $+$ \batse\ 
time-resolved spectra of 
GRBs 970111, 980329, 990123 and 990510 are reported in 
Table~\ref{t:bf}. 
The time behavior of the best-fit parameters 
to the tested model is  shown in Figure~\ref{f:time_evol}. 
In the bottom panels of this figure, the time 
behavior of the null hypothesis probability (NHP) is also shown. 

From Figure~\ref{f:time_evol} it is apparent that $E_p$, for each event, mimics the time 
behavior of the 2--2000 keV flux. This result is better shown in 
Figure~\ref{f:ep-vs-flux}, where the time-resolved peak energy derived from the 
best fit of BF to the time-resolved spectra is plotted as a function of the 
corresponding flux. 
As can be seen from this figure, a positive correlation is found for each of the
four GRBs, consistent with a power-law dependence of the peak energy on flux, even if the 
statistical significance changes from one event to another, also depending 
on the number of available data points (see Table~\ref{t:ep-vs-flux}). 
The power-law parameters were derived in two ways: using the simple least-squares method, 
and the maximum 
likelihood method in the case that the correlated data ($x_i$,$y_i$) could be described 
by a linear function
$Y = mX + q$ with the addition of an extrinsic (non Poissonian) variance $\sigma^2_{\rm ext}$
among the free parameters \citep{Dagostini05}. The latter method has already been adopted for
various applications (see, e.g., \citet{Amati08} and references therein).

Given that the peak energy $E_p$ is related to both $E_0$ and $\alpha$
($E_p = E_0 (2+ \alpha)$), in order to establish the $\alpha$ contribution to the
strong correlation between  $E_p$ and flux, we investigated the behavior of $\alpha$
with $E_p$. The result is shown in Figure~\ref{f:alpha-vs-ep}. As can be seen, even if,  in 
two cases
(GRB\,970111 and GRB\,980329), we see a rapid change of $\alpha$ with $E_p$, no 
significant correlation is found
in the cases of the other GRBs, especially in the case of GRB\,990123, where we have the
highest number of time-resolved spectra. 

For GRBs with known redshift (GRB\,990123 and GRB\,990510), using the concordance cosmology 
($\Omega_\lambda = 0.73$, $\Omega_m = 0.27$, 
$H_0 = 70$~km~s$^{-1}$~Mpc$^{-1}$), we derived the intrinsic 
time resolved peak energy $E_{p,i}$ as a function of the corresponding 2--2000 keV 
isotropic luminosity $L_{\rm iso}$. The result is shown in Figure~\ref{f:Epi-vs-L}. It does
not change if we integrate $L_{\rm iso}$ up to 10 MeV using the best fit parameters of the BF.
As can be seen from this figure, and quantified by the best fit correlation results 
(see Table~\ref{t:epi-vs-L}), $E_{p,i}$ is related to 
$L_{\rm iso}$ through a power-law, 
with a very low probability that the result is due to chance, especially in the case of GRB\,990123. 
For each of these GRBs,  we also derived  $L_{\rm iso}$ separately in the energy band below 
$E_{p,i}$  and above 
$E_{p,i}$. The result is that in both ranges the correlation between $E_{p,i}$ 
and $L_{\rm iso}$ still holds. 

In order to see the cumulative result, we have merged all the available data on GRBs 990123 
and 990510, to derive the mean dependence of $E_{p,i}$ on $L_{\rm iso}$. The result is shown in  
Figure~\ref{f:epi-all-vs-L}. 
A power-law dependence of $E_{p,i}$ on $L_{\rm iso}$ is still found 
(see Table~\ref{t:epi-vs-L}), with a very low null hypothesis probability ($1.57\times10^{-13}$),
much lower than that found within 
each event, confirming the reality of the $E_p$--$L_{\rm iso}$ correlation in the single events.

\section{Discussion and conclusions}

By joining together the WFC and \batse\ spectral data  of the four strongest 
($>15\times 10^{-6}$~erg~cm$^{-2}$) GRBs simultaneously observed  with 
both instruments, it was possible to perform a fine 
time-resolved spectral analysis in 
the broad energy band 2 keV to 2 MeV, a passband still scarcely explored as a whole. 

We do not expect significant systematic errors from this joint analysis. 
The response functions of both instruments are well known. Indeed, in the fits 
the cross-calibration factor was found to be always 1, in spite of being left to 
vary between 0.8 and 1.2 (see Section~\ref{analysis}). This is not the first time that
a joint WFC/\batse\ spectral analysis has been performed. Results of similar analyses have been 
reported in the past by the \batse\ team \citep{Briggs00,Kippen03,Kippen04}. In addition, the 
\batse--deconvolved spectra of bright GRBs were cross-checked with those obtained with
\sax\ GRBM \citep{Frontera09}; these in turn were cross-calibrated with WFC, 
with many published results \citep[e.g.][]{Frontera98a,Frontera00a}.

For each of the strongest GRBs, we obtained a significant 
number of time-resolved spectra with constrained $E_p$: 8 spectra for GRB\,970111, 6 spectra 
for GRB\,980329, 19  spectra for GRB\,990123, and 7 spectra for GRB\,990510, with a 
total number of 40 analyzed spectra. 
With these spectra, we investigated the dependence of 
the time-resolved peak energy $E_p$ on the corresponding 
2--2000 keV flux using the empirical BF as an input model.

We find a significant power-law correlation between the derived 
peak energy $E_p$ and the flux within each GRB in our sample. 
The power-law best-fit parameters,  evaluated with two different methods 
(the least-squares method, and the likelihood method 
with the addition of an external variance as a free parameter) give 
(see Table~\ref{t:ep-vs-flux}) similar index values in the cases of  
GRBs 970111 ($0.68\pm 0.06$ versus $0.65_{-0.14}^{+0.16}$)  and 
980329 ($0.16\pm 0.04$ versus $0.15_{-0.07}^{+0.08}$), and different values 
in the case of the GRBs 990123 ($0.53\pm 0.05$ versus $0.46_{-0.09}^{+0.09}$) 
and 990510 ($0.81\pm 0.15$ versus $0.56_{-0.23}^{+0.25}$), 
even if, in the latter case, these values are statistically almost consistent 
with each other. However, the likelihood method has the advantage of giving us 
information about the non-Poissonian spread of the data points 
around the best-fit curve. It shows that such a spread, even if small, affects the correlation, 
with the highest value  ($\sigma_{\rm ext} = 0.09$~dex) in the case of GRB\,990510 and 
the minimum one ($\sigma_{\rm ext} = 0.00_{-0.00}^{+0.04}$~dex) in the case of GRB\,980329. 
Actually, in this case, also due to the low statistics of the data points, 
the correlation significance is low and, as shown by the reduced $\chi^2$ value 
(see Table~\ref{t:ep-vs-flux}), is not sensitive to an external spread. 

No clear correlation is found  between the low-energy photon index $\alpha$ and flux and 
between $\alpha$ and $E_p$, apart from in one case (GRB\,980329). This means that the most relevant parameter that gives rise to
the $E_p$--flux correlation is the break parameter $E_0$ in the BF.

An equivalent important result is that the power-law correlation found between $E_p$ and flux 
is confirmed when we correlate, for two GRBs with known redshift,  the intrinsic 
peak energy $E_{p,i}$ with the corresponding isotropic luminosity. 
Notice that the power-law index of the correlation changes from one GRB to 
the next when we use the least-squares method ($0.53\pm 0.05$ 
for GRB\,990123, and $0.81\pm 0.15$ for GRB\, 990510), but
does not  when the likelihood method is used ($0.46 \pm 0.09$ for GRB\,990123, $0.56_{-0.23}^{+0.25}$ for GRB\, 990510). This fact clearly means that the 
power-law index is affected by the external spread. If this spread is due to an 
unknown physical parameter, this result is an important hint for physical models of 
the prompt emission process: The  intrinsic $E_{p,i}$ derived from the assumed emission 
model should be related to at least two physical parameters of the model.
 
When we join together all the available data on GRBs with known redshift
we find that the $E_{p,i}$--$L_{\rm iso}$ correlation becomes even 
more robust, with a probability of $1.57\times 10^{-13}$ that the  correlation averaged 
over all data is due to chance, and a power-law index that is almost independent of the 
used best fit method ($0.66\pm 0.03$ in the case of the 
least-square method versus 
$0.63_{-0.07}^{+0.06}$ in the case of the likelihood method). 
It is also interesting to note that the non-Poissonian spread found for the GRB 
averaged correlation
($\sigma_{\rm ext} = 0.06_{-0.05}^{+0.06}$) is three times lower than the spread found in the 
Amati relation \citep{Amati09}, which is well known to be based on  time-averaged 
spectra. This is   a strong hint that part of the spread of the Amati relation is related 
with the fact that the $E_{p}$ values determined from 
time-averaged spectra are 
biased because of the spectral evolution of the GRB prompt emission. 
 
If we compare our GRB-averaged correlation result with that obtained by \citet{Ghirlanda10} 
from the time-resolved spectra of \fermi\ GRBs, we find that
our results are consistent with those  within a 2$\sigma$ belt, even if the slope 
obtained by these authors ($0.36 \pm 0.05$) is lower than that found by us 
($0.63_{-0.07}^{+0.06}$; see Table~\ref{t:epi-vs-L} and
 Figure~\ref{f:epi-all-vs-L}), likely due to the sample variance. Indeed,
our found slope is similar to that ($0.621 \pm 0.003$) reported by \citet{Lu12}, who performed
the time-resolved spectral analysis of a sample of 15 \fermi\ GRBs with known redshift. 
 
Within a $2 \sigma$ spread, our results are also
consistent with the time-averaged $E_{p,i}$ versus $L_{\rm iso}$ correlation, 
obtained by \citet{Ghirlanda10} using 95 pre-\fermi\ plus 10 \fermi\ GRBs 
(see the right panel of Figure~\ref{f:ep-vs-L-comp}). 

We cannot exclude that selection effects can influence our results, as with the fact that
the analyzed GRBs are detected by both instruments, that they are sufficiently bright to allow
time resolved analysis, and that they have known redshift. However, it seems difficult that 
this unavoidable selection can introduce a correlation between $E_p$ and $L_{\rm iso}$ within single GRBs. In addition, a similar 
correlation has been found by the other mentioned authors \citep{Ghirlanda10, Lu12} with
other GRBs and with other instruments. 

To conclude our results strongly support, at least in the range of luminosities 
explored with our data, the reality of the Amati  \citep{Amati02} and the Yonetoku 
\citep{Yonetoku04} relations, both derived using time averaged spectra. 

Also,  our results give strong constraints on the physical models. In a forthocoming paper (Frontera et al. 2012, in preparation), with the same data we are testing different physical 
models, among them the recently developed {\sc grbcomp} model,
which is devoted to the spectral formation of a GRB \citep{Titarchuk12}.  In this model a physical interpretation of the Amati relation is also given.

\begin{acknowledgments}
We are grateful to Pawan Kumar for valuable discussions and suggestions. The \sax\ satellite
was a joint effort of the Italian Space Agency and the Netherland Space Agency. This research made use
of data obtained through the HEASARC Online service provided by the NASA 
Goddard Space Flight Center. We acknowledge its support.

\end{acknowledgments}

\bibliographystyle{apj}
\bibliography{grb_ref}

\clearpage

%
%
\begin{deluxetable}{lcccc}
\tabletypesize{\small}
\tablewidth{0pt}
\tablecaption{GRB Sample Chosen for the Time-Resolved Spectral Analysis}
\tablehead{
\colhead{GRB} & \colhead{Redshift} & \colhead{Fluence} &
\colhead{No. of Intervals} & \colhead{No. of Useful Intervals} \\
 & &  \colhead{($\times 10^{-6}$~cgs)} &	& 
} 
\startdata
970111 & -- & 39.18$\pm$0.08  & 10 & 8 \\
980329 & -- & $37.53\pm0.07$ & 8 & 6  \\
990123 & 1.60 &	$205.12\pm0.03$ & 26 & 19 \\
990510 & 1.619 & $15.80\pm0.07$ & 15 & 8  \\
\enddata
\tablecomments{ 
For each GRB we report the redshift, the fluence in the 2-2000 keV energy band, the 
number of time intervals in which we subdivided the time profile and those used in the time-resolved spectral analysis.
}
\label{t:grb-intervals}
\end{deluxetable}		

\clearpage
%
%
\begin{deluxetable}{lccccccl}
\tabletypesize{\small}
\tablewidth{0pt}
\tablecaption{Best fit Parameters of the BF in the Time Intervals
for Which a Sensitive Spectral Analysis Was Possible}
\tablehead{ 
\colhead{GRB} & \colhead{Interval} & \colhead{Start ($\Delta$t)} & \colhead{$\alpha$} & \colhead{$\beta$}  & \colhead{$E_p$} & 
\colhead{$\chi^2/{\rm dof}$} & \colhead{Flux}  \\  
 &  & \colhead{(SOD)} & & & \colhead{(keV)} &  & \colhead{($\times 10^{-6}$~cgs)} \\
} 
\startdata
970111	 & 3  & 35045.0 (3) & $0.57_{-0.06}^{+0.08}$  &  $-3.99_{-0.43}^{+0.32}$  & $203_{-7}^{+8}$ & 114.2/97 & $1.98\pm0.55$ \\
	 & 4  & 35048.0 (6) & $-0.23_{-0.02}^{+0.06}$ &  $-4.89_{-5.11}^{+0.77}$\tablenotemark{a}  & $174_{-8}^{+3}$ & 87.2/75 &  $1.79\pm0.20$  \\
	 & 5  & 35054.0 (3) & $-0.32_{-0.03}^{+0.05}$ &  $-9.37_{-0.63}^{+4.35}$\tablenotemark{a}  & $162_{-2}^{+5}$ & 89.2/73 &  $2.39\pm0.18$  \\
	 & 6  & 35057.0 (5) & $-0.48_{-0.03}^{+0.03}$ &  $-4.94_{-5.06}^{+0.56}$\tablenotemark{a}  & $151_{-3}^{+4}$ & 83.0/73 &  $2.44\pm0.16$  \\
	 & 7  & 35062.0 (3) & $-0.58_{-0.04}^{+0.04}$ &  $-3.46_{-0.21}^{+0.15}$  & $89_{-3}^{+4}$ & 29.3/13  &  $1.02\pm0.14$  \\
	 & 8  & 35065.0 (4) & $-0.58_{-0.03}^{+0.02}$ &  $-3.74_{-0.23}^{+0.21}$  & $64_{-2}^{+2}$ & 32.3/13  &  $0.58\pm0.07$  \\
	 & 9  & 35069.0 (4) & $-0.61_{-0.06}^{+0.08}$ &  $-3.15_{-0.09}^{+0.09}$  & $56_{-6}^{+5}$ & 15.6/12  &  $0.71\pm0.38$  \\
	 & 10 & 35073.0 (13) & $-0.69_{-0.05}^{+0.04}$ &  $-2.85_{-0.07}^{+0.08}$  & $57_{-3}^{+5}$ & 22.4/13  &  $0.43\pm0.15$  \\
	&		&	 	 &		 &		&    	 \\
980329   & 2  & 13477.0 (4) & $-0.74_{-0.07}^{+0.09}$ &  $-2.17_{-0.12}^{+0.10}$  & $256_{-32}^{+35}$ & 105.7/83   &  $1.50\pm0.49$  \\
	 & 3  & 13481.0 (2) & $-0.61_{-0.05}^{+0.06}$ &  $-2.46_{-0.10}^{+0.08}$  & $236_{-9}^{+19}$ & 108.1/90    &  $3.25\pm0.66$  \\
	 & 4  & 13483.0 (3) & $-0.70_{-0.05}^{+0.03}$ &  $-2.49_{-0.05}^{+0.06}$  & $235_{-9}^{+16}$ & 105.9/97    &  $4.82\pm0.70$  \\
	 & 5  & 13486.0 (4) & $-0.80_{-0.02}^{+0.05}$ &  $-2.22_{-0.04}^{+0.04}$  & $242_{-17}^{+8}$ & 135.5/105   &  $4.25\pm0.41$  \\
	 & 6  & 13490.0 (4) & $-1.21_{-0.05}^{+0.05}$ &  $-2.22_{-0.09}^{+0.07}$  & $169_{-19}^{+21}$ & 97.7/77    &  $0.95\pm0.15$  \\
	 & 7  & 13494.0 (19) & $-1.49_{-0.05}^{+0.07}$ & $-2.79_{-7.21}^{+0.46}$\tablenotemark{a}  & $136_{-23}^{+32}$ & 52.8/47                   &  $0.15\pm0.04$  \\
 	 &		&	 	 &		 &		&    	 \\
990123	 & 2  & 35221.9 (6) & $-0.44_{-0.11}^{+0.13}$ &  $-9.37_{-0.63}^{+6.56}$\tablenotemark{a}   & $148_{-34}^{+32}$ & 39.2/35                      &  $0.22\pm0.13$ \\
	 & 3  & 35227.9 (6) & $-0.19_{-0.32}^{+0.10}$ &  $-2.52_{-7.48}^{+0.40}$ \tablenotemark{a}   & $171_{-25}^{+97}$ & 27.8/44                      &  $0.64\pm0.49$ \\
	 & 4  & 35233.9 (2) & $-0.58_{-0.11}^{+0.12}$ &  $-9.30_{-0.70}^{+6.70}$\tablenotemark{a}   & $388_{-49}^{+96}$ & 55.1/57                      &  $1.71\pm0.90$ \\
	 & 5  & 35235.9 (2) & $-0.55_{-0.03}^{+0.03}$ &  $-9.37_{-0.63}^{+6.96}$\tablenotemark{a}   & $598_{-53}^{+52}$ & 101.9/105                    &  $3.80\pm0.83$  \\
	 & 6  & 35237.9 (2) & $-0.56_{-0.08}^{+0.03}$ &  $-2.07_{-7.92}^{+0.39}$\tablenotemark{a}   & $844_{-145}^{+251}$ & 152.8/151                  &  $8.64\pm3.01$  \\
	 & 7  & 35239.9 (2) & $-0.55_{-0.04}^{+0.03}$ &  $-2.12_{-0.71}^{+0.19}$  & $979_{-97}^{+149}$ & 177.3/182     &  $13.87\pm2.60$ \\
	 & 8  & 35241.9 (2) & $-0.54_{-0.04}^{+0.02}$ &  $-2.29_{-7.72}^{+0.28}$\tablenotemark{a}   & $1094_{-108}^{+220}$ & 164.7/183                 &  $14.70\pm2.91$ \\
	 & 9  & 35243.9 (2) & $-0.51_{-0.06}^{+0.04}$ &  $-1.98_{-0.37}^{+0.13}$  & $594_{-64}^{+130}$ & 192.0/166     &  $8.58\pm2.41$  \\
	 & 10 & 35245.9 (2) & $-0.56_{-0.03}^{+0.02}$ &  $-9.37_{-0.62}^{+5.36}$\tablenotemark{a}   & $478_{-22}^{+28}$ & 149.3/140                    &  $5.19\pm0.68$  \\
	 & 11 & 35247.9 (2) & $-0.66_{-0.04}^{+0.07}$ &  $-3.66_{-6.33}^{+1.10}$\tablenotemark{a}   & $272_{-40}^{+30}$ & 98.9/104                     &  $2.12\pm0.61$  \\
	 & 12 & 35249.9 (2) & $-0.51_{-0.04}^{+0.05}$ &  $-9.36_{-0.63}^{+6.04}$\tablenotemark{a}   & $431_{-29}^{+39}$ & 136.5/118                    &  $4.35\pm0.81$  \\
	 & 13 & 35251.9 (2) & $-0.44_{-0.04}^{+0.05}$ &  $-2.28_{-0.29}^{+0.18}$  & $582_{-64}^{+65}$ & 211.4/188      &  $9.53\pm2.11$  \\
	 & 14 & 35253.9 (2) & $-0.58_{-0.03}^{+0.04}$ &  $-2.27_{-0.35}^{+0.20}$  & $712_{-69}^{+63}$ & 200.0/180      &  $9.84\pm1.67$  \\
	 & 15 & 35255.9 (2) & $-0.61_{-0.03}^{+0.04}$ &  $-2.44_{-0.30}^{+0.25}$  & $610_{-69}^{+41}$ & 169.1/182      &  $7.54\pm1.26$  \\
	 & 16 & 35257.9 (2) & $-0.71_{-0.04}^{+0.04}$ &  $-2.77_{-7.23}^{+0.40}$\tablenotemark{a}   & $478_{-52}^{+60}$ & 140.6/125                    &  $5.13\pm1.09$  \\
	 & 17 & 35259.9 (5) & $-0.91_{-0.03}^{+0.02}$ &  $-3.29_{-6.71}^{+0.67}$\tablenotemark{a}   & $414_{-34}^{+34}$ & 160.5/170                    &  $3.02\pm0.42$  \\
	 & 18 & 35264.9 (5) & $-1.01_{-0.03}^{+0.03}$ &  $-2.00_{-0.19}^{+0.12}$  & $260_{-37}^{+50}$ & 164.4/147      &  $2.16\pm0.42$  \\
	 & 19 & 35269.9 (5) & $-1.08_{-0.03}^{+0.03}$ &  $-2.62_{-7.37}^{+0.56}$\tablenotemark{a}   & $361_{-55}^{+57}$ & 140.8/147                    &  $1.96\pm0.37$  \\
	 & 20 & 35274.9 (5) & $-1.13_{-0.03}^{+0.01}$ &  $-9.36_{-0.64}^{+6.79}$\tablenotemark{a}   & $406_{-46}^{+70}$ & 112.9/134                    &  $1.73\pm0.28$  \\
       	&		&	 	 &		 &		&    	 \\
990510   & 1  & 31745.9 (5) & $-0.67_{-0.09}^{+0.09}$ &  $-2.89_{-0.28}^{+0.15}$  & $80_{-10}^{+9}$   & 73.9/48        &  $0.42\pm0.16$  \\
	 & 2  & 31750.9 (5) & $-0.96_{-0.07}^{+0.12}$ &  $-3.55_{-1.63}^{+0.33}$  & $86_{-9}^{+7}$    & 53.4/53        &  $0.54\pm0.20$  \\
	 & 6  & 31770.9 (15) & $-0.97_{-0.06}^{+0.07}$ &  $-2.60_{-0.38}^{+0.22}$  & $264_{-34}^{+34}$ & 95.3/81        &  $1.29\pm0.40$  \\
	 & 7  & 31785.9 (5) & $-0.80_{-0.06}^{+0.08}$ &  $-2.57_{-0.11}^{+0.10}$  & $178_{-17}^{+14}$ & 100.3/80       &  $2.05\pm0.63$  \\
	 & 8  & 31790.9 (3) & $-1.14_{-0.04}^{+0.03}$ &  $-9.37_{-0.63}^{+6.31}$\tablenotemark{a}   & $151_{-11}^{+13}$ & 70.9/62                      &  $0.63\pm0.13$  \\
	 & 9  & 31793.9 (4) & $-1.01_{-0.35}^{+0.18}$ &  $-2.35_{-0.06}^{+0.05}$  & $31_{-10}^{+71}$  & 28.7/39        &  $1.07\pm1.03$  \\
	 & 10 & 31797.9 (5) & $-1.14_{-0.12}^{+0.05}$ &  $-2.86_{-0.12}^{+0.11}$  & $64_{-5}^{+2}$    & 20.0/16        &  $0.45\pm0.13$  \\
	 & 11 & 31802.9 (4) & $-1.70_{-0.05}^{+0.10}$ &  $-9.37_{-0.63}^{+6.58}$\tablenotemark{a}  & $49_{-11}^{+11}$  & 40.9/34                       &  $0.11\pm0.04$  \\
\enddata\label{t:bf}
\tablecomments{
Uncertainties are single parameter errors at 90\% confidence level. For each time interval of Figure~\ref{f:lc},  we also report the Seconds Of Day (SOD) in correspondence of its start, its duration $\Delta t$ in seconds, and the corresponding flux in the 2--2000 keV energy
band. For the flux estimate, in the cases  the high energy index $\beta$ could not be constrained, it was frozen to $-2.3$.
}
\tablenotetext{a}{Value fixed to $-2.3$ in the computation of flux and luminosity.}
\end{deluxetable}

%
%
\begin{deluxetable}{lcccccl}
\tablewidth{0pt}
\tablecaption{Correlation Analysis Results between $E_p$ and 2--2000 keV Flux}
\tablehead{
\colhead{GRB} & \colhead{$\rho$} & \colhead{NHP} 
& \colhead{$k$} & \colhead{$m$} & \colhead{$\sigma_{\rm ext}$}  &  \colhead{$\chi_r^2$ (dof)}  \\  
} 
\startdata
970111	& 0.69 & $5.0 \times 10^{-2}$    & $6.05\pm 0.34$  	    & $0.68 \pm 0.06$        &      			& 2.9 (6) \\
		&	 &				   & $5.91_{-0.78}^{+0.91}$ & $0.65_{-0.14}^{+0.16}$ & $0.04_{-0.03}^{+0.05}$ & 1.5 (5) \\
980329	& 0.49 & 0.32  			   & $3.25\pm 0.26$ 	    & $0.16 \pm 0.04$ 		& 				& 1.2 (4) \\
		&	 &				   & $3.20_{-0.40}^{+0.43}$ & $0.15_{-0.07}^{+0.08}$ & $0.00_{-0.00}^{+0.04}$ & 1.6 (3) \\
990123	& 0.92 & $2.68 \times 10^{-8}$   & $5.51\pm 0.28$ 	    & $0.53 \pm 0.05$ 	     & 				& 1.3 (17)	\\
		&	 &				   & $5.16_{-0.48}^{+0.49}$ & $0.46_{-0.09}^{+0.09}$ & $0.04_{-0.04}^{+0.11}$ & 1.1 (16) \\
990510	& 0.88 & $3.85 \times 10^{-3}$   & $7.09\pm 1.01$ 		& $0.81 \pm 0.15$ 	&				& 1.8 (6)	\\
		&	 &				   & $5.54_{-1.43}^{+1.55}$ & $0.56_{-0.23}^{+0.25}$ & $0.09_{-0.09}^{+0.10}$ & 1.4 (5) \\

\enddata
\tablecomments{
A power-law relation between the two parameters is assumed: $\log{E_p} = k + m \log{flux}$. 
In addition to the best fit parameters $k$ and $m$, reported are
the Spearman correlation coefficient $\rho$, NHP, and the best fit $\chi_r^2$ with the degrees of freedom dof.
The parameter estimate was performed using both the least-squares method and the likelihood method, in which case
the extrinsic variance $\sigma_{\rm ext}$ between the $E_{p,i}$ and flux is included in the log-likelihood function 
as a free parameter(see the text).
}

\label{t:ep-vs-flux}
\end{deluxetable}

%
%
\begin{deluxetable}{lcccccl}
\tablewidth{0pt}
\tablecaption{Correlation Analysis Results between the Intrinsic (Redshift Corrected) Peak Energy $E_{p,i}$
and the GRB Luminosity, Either for Individual GRBs with Known Redshift (990123, 990510) or When All the Data Points of These GRBs Are Taken Into Account}  
\tablehead{
\colhead{GRB} & \colhead{$\rho$} & \colhead{NHP} & \colhead{$k$} & \colhead{$m$} & \colhead{$\sigma_{\rm ext}$} & \colhead{$\chi_r^2$ (dof)} \\    
} 
\startdata
990123	& 0.92	& $2.7\times 10^{-8}$ & $2.64\pm 0.05$  		& $0.53\pm 0.05$ 		& 				&  1.3 (17) \\
		&		&			    & $2.71_{-0.09}^{+0.09}$ & $0.46_{-0.09}^{+0.09}$ & $0.04_{-0.04}^{+0.11}$ & 1.1 (16) \\
990510	& 0.88 	& $3.85\times 10^{-3}$ & $2.46\pm 0.04$ 		& $0.81\pm 0.15$ 		& 				& 1.8  (6) \\
		&		&			    & $2.45_{-0.09}^{+0.08}$ & $0.56_{-0.23}^{+0.25}$ & $0.09_{-0.09}^{+0.10}$ & 1.4 (5) \\
full sample & 0.94  	& $1.57 \times 10^{-13}$ & $2.51\pm 0.03$  	& $0.66 \pm 0.03$ 	& 				& 1.6 (25) \\
		&		&			    & $2.53_{-0.06}^{+0.06}$ & $0.63_{-0.07}^{+0.06}$ & $0.06_{-0.06}^{+0.05}$ & 1.2 (24) \\
\enddata
\tablecomments{
A power-law relation between the two parameters is assumed: $\log{E_{p,i}} = k + m \log{L_{iso}}$. 
In addition to the best fit parameters $k$ and $m$, reported are
the Spearman correlation coefficient $\rho$, NHP, and the best fit reduced $\chi_r^2$ with  dof.
The parameter estimate was performed using both the least-squares method and the likelihood method, in which case
the extrinsic variance $\sigma_{\rm ext}$ between the $E_{p,i}$ and flux is included in the log-likelihood function 
as a free parameter(see the text).
}
\label{t:epi-vs-L}
\end{deluxetable}

\clearpage

%
%
\begin{figure}
\begin{center}
  \includegraphics[width=.4\textwidth]{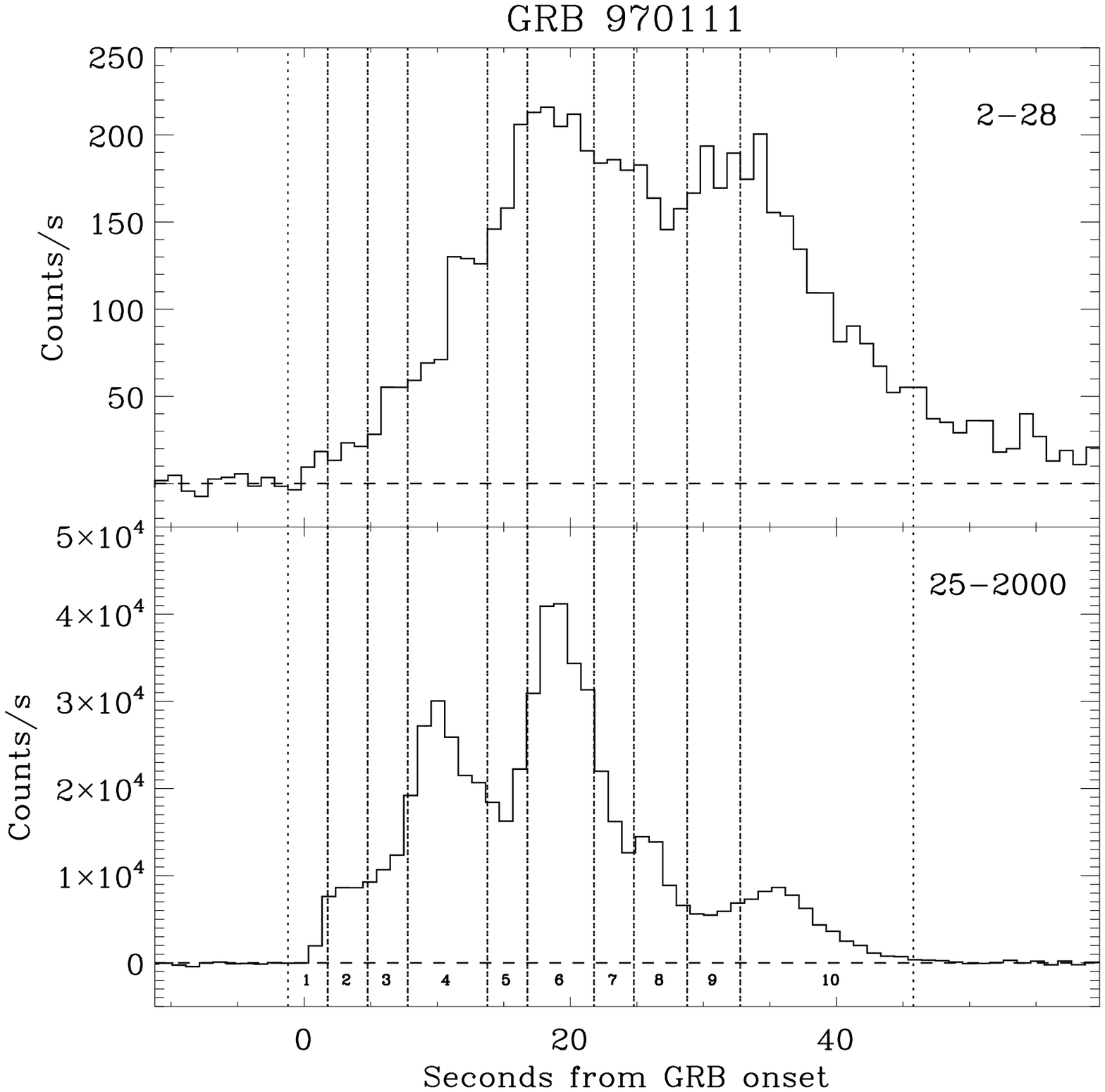}
  \includegraphics[width=.4\textwidth]{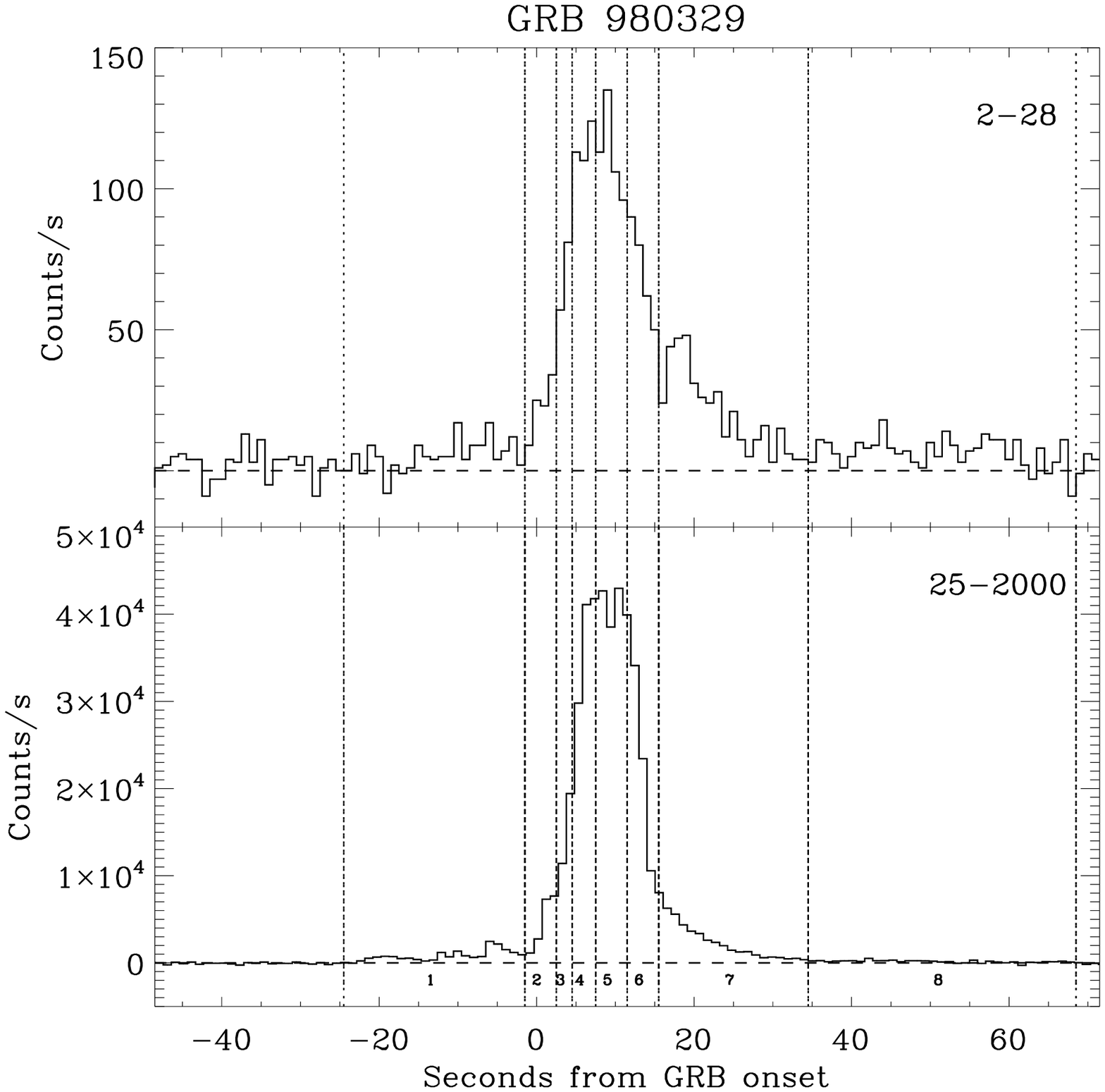}

\vspace{0.5in}

  \includegraphics[width=.4\textwidth]{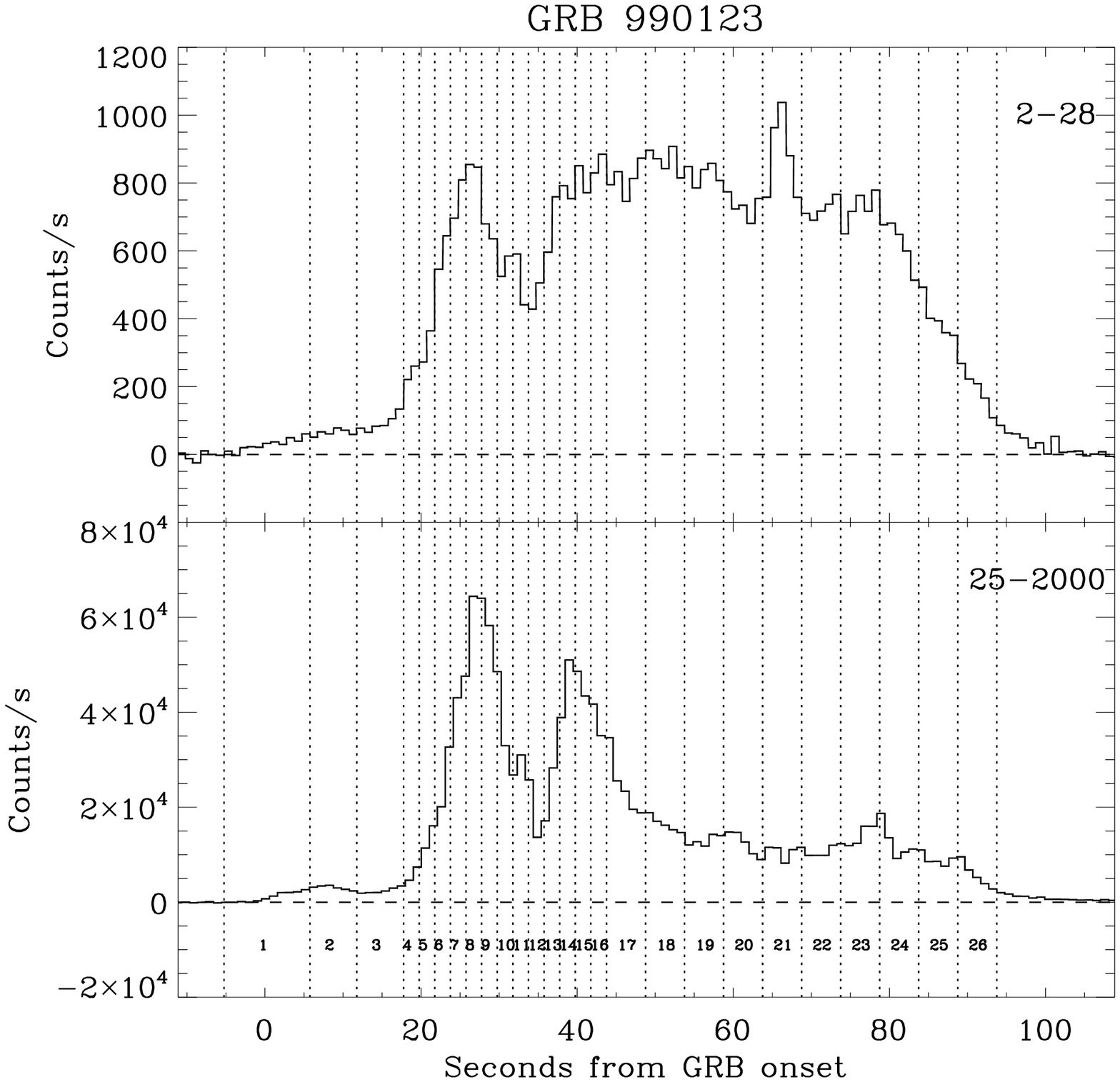}
  \includegraphics[width=.4\textwidth]{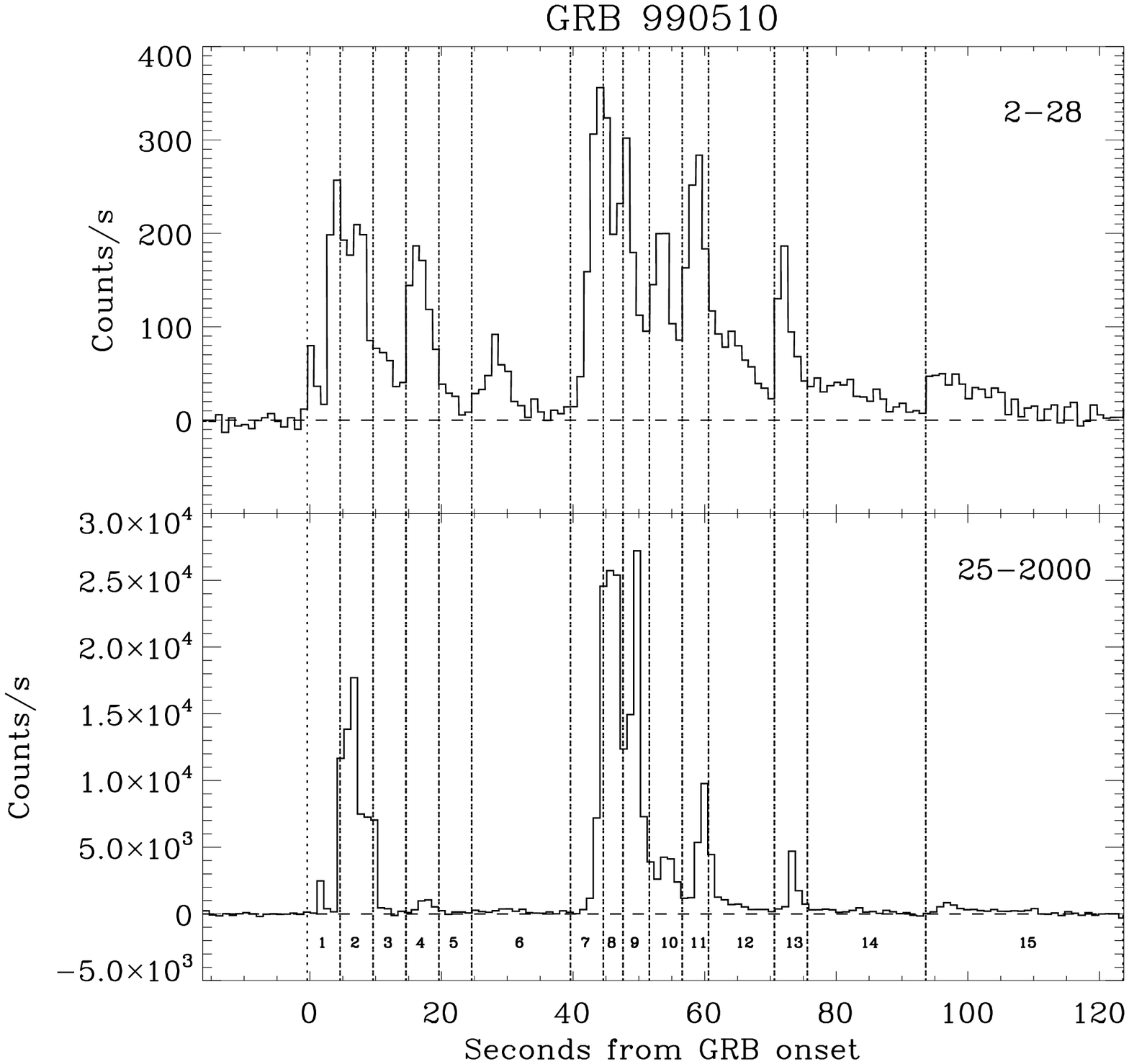}
\end{center}
\vspace{0.5cm}
\caption{Light curves of GRBs 970111, 980329, 990123, and 990510, detected with
\sax\ WFC (2--28 keV) and \batse\ (25--2000 keV). Also shown are the intervals in which the time
resolved spectra were derived.}
\label{f:lc}
\end{figure}

\clearpage

%
%
\begin{figure}
\begin{center}
  \includegraphics[width=.4\textwidth]{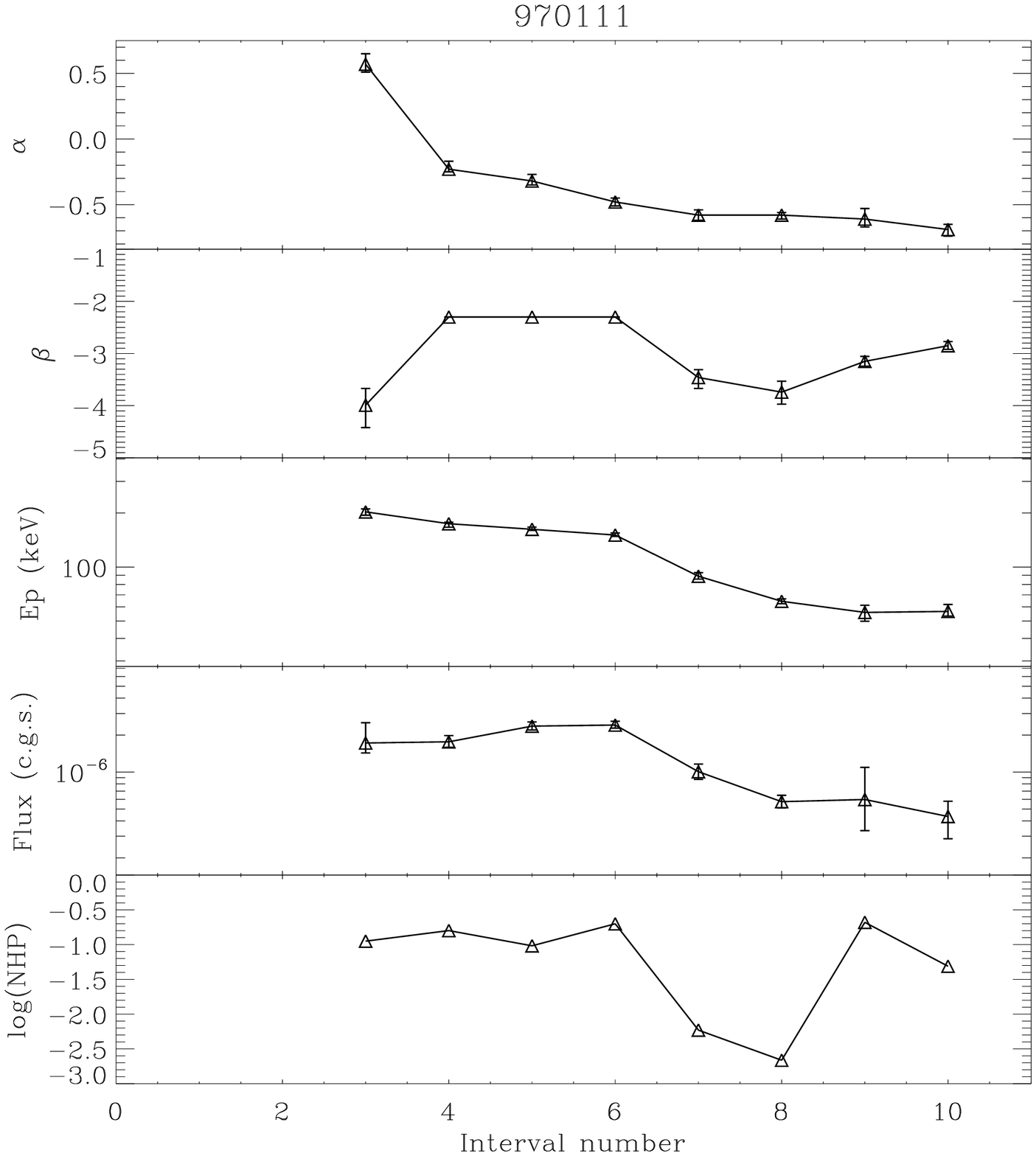}
  \includegraphics[width=.4\textwidth]{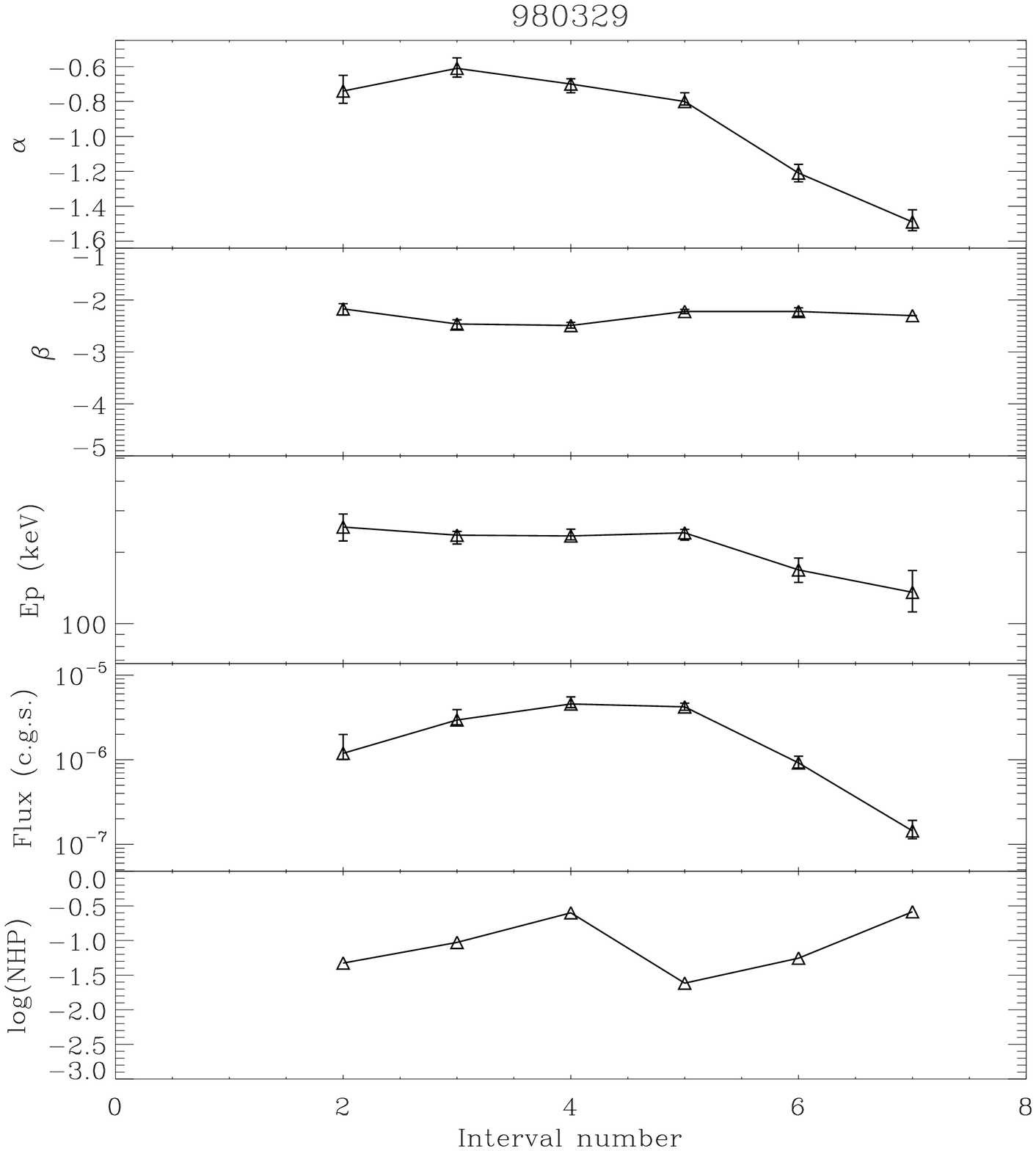}

\vspace{0.5in}

  \includegraphics[width=.4\textwidth]{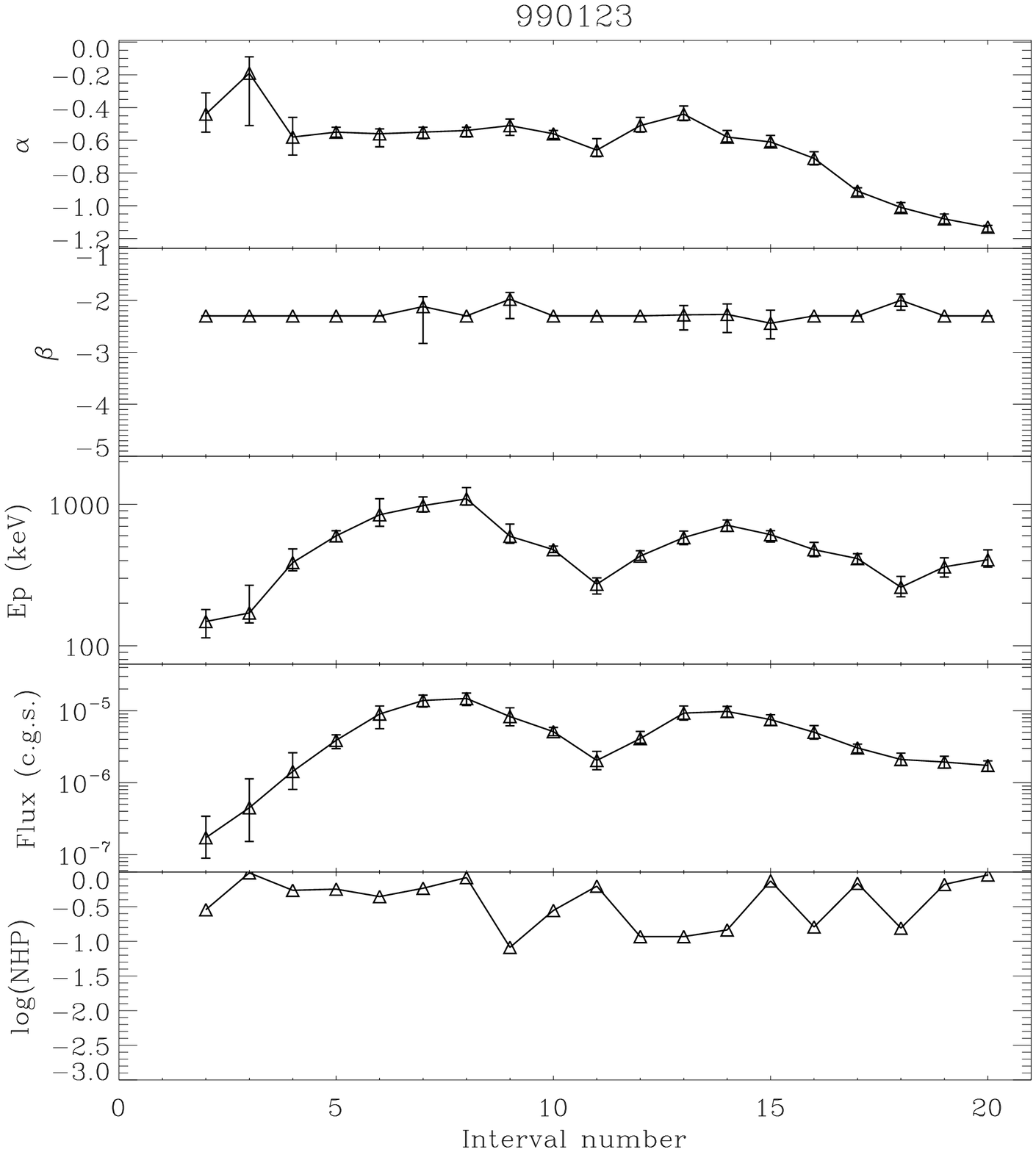}
  \includegraphics[width=.4\textwidth]{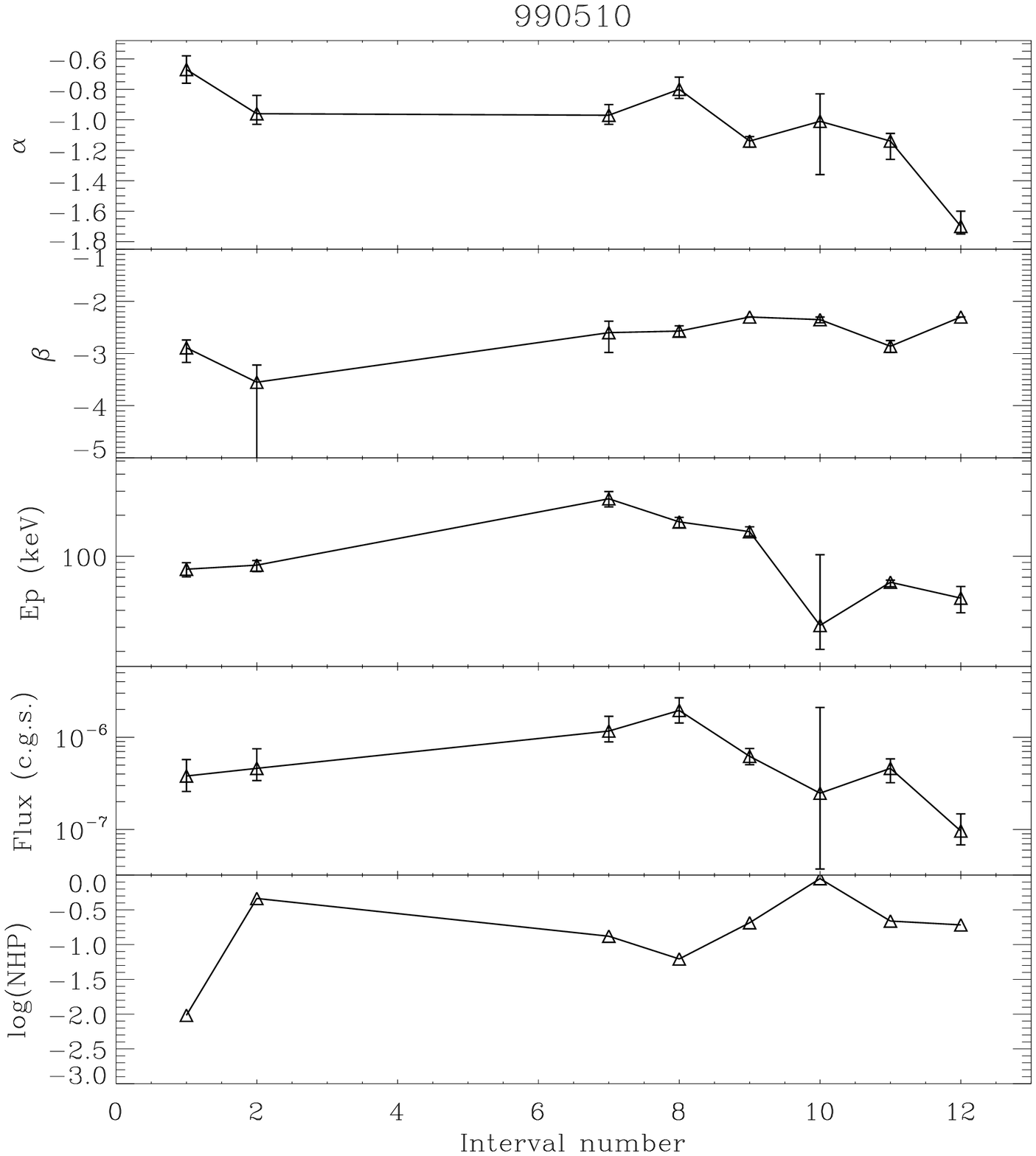}
\end{center}
  \caption{Temporal evolution of the BF best fit parameters and of 
the null hypothesis probability (NHP),for each of the GRBs 970111, 980329, 990123, 990510. 
Top panel: low energy {\sc pl} index $\alpha$. 
Second panel from the top: high energy power-law index $\beta$, in the cases in which it
was constrained from the fit. In the other cases (not plotted) a value of $\beta= -2.3$
is assumed. 
Third panel from the top: peak energy $E_p$.
Fourth panel from the top: estimated  2--2000 keV flux.
Bottom panel: null hypothesis probability (NHP) for 
the tested function.} 
\label{f:time_evol}
\end{figure}

\clearpage

%
%
\begin{figure}
\begin{center}
  \includegraphics[width=.4\textwidth]{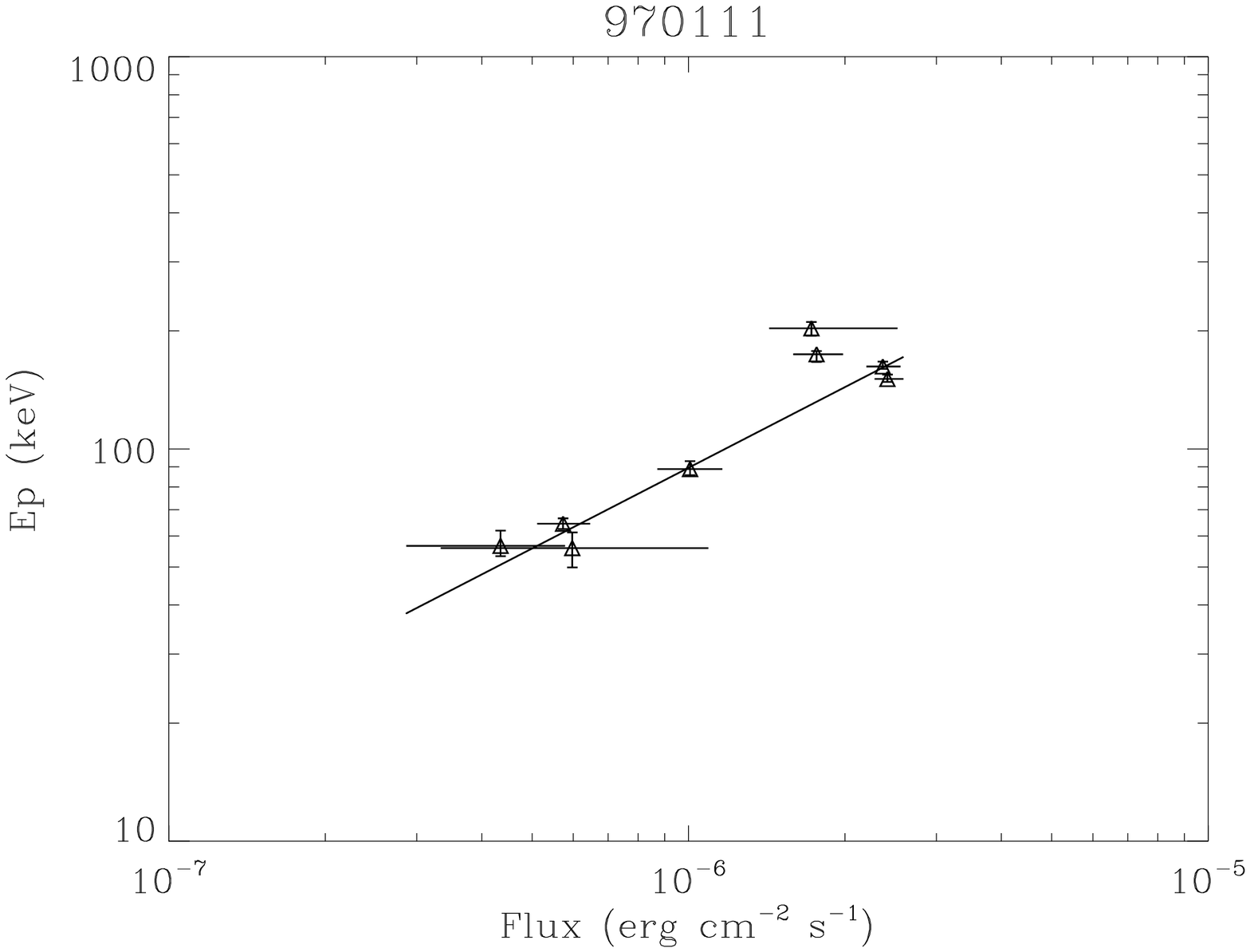}
  \includegraphics[width=.4\textwidth]{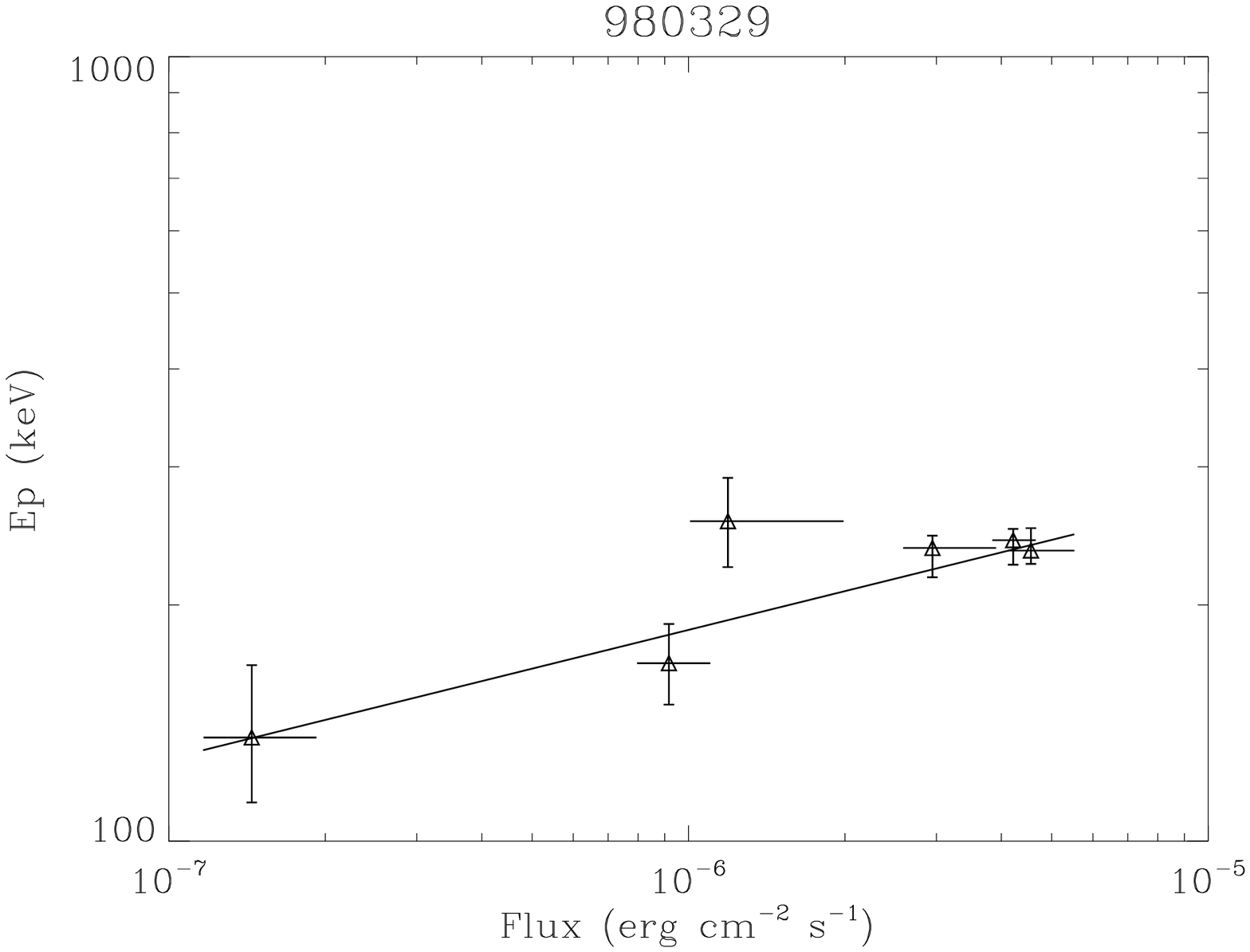}
  \includegraphics[width=.4\textwidth]{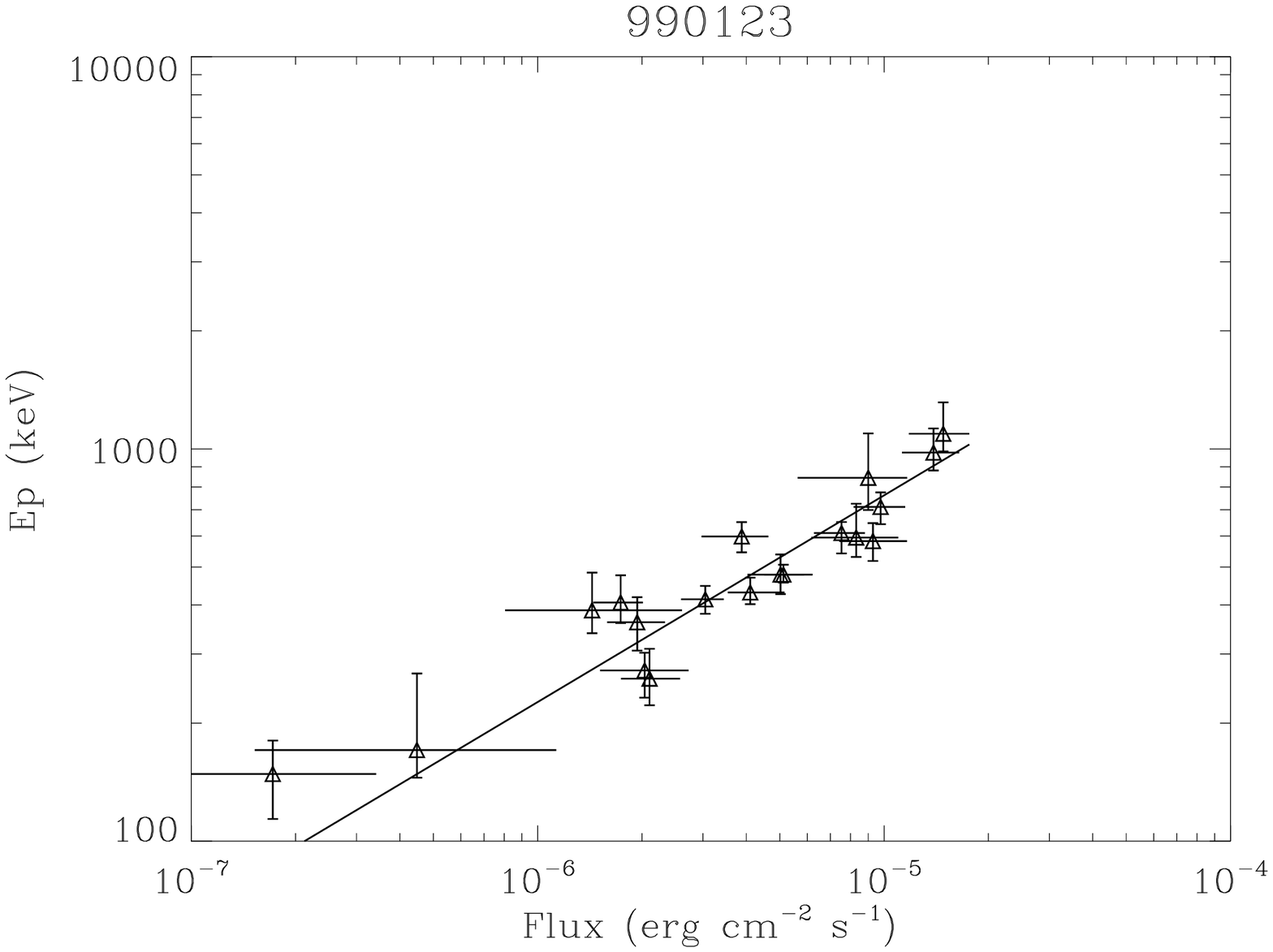}
  \includegraphics[width=.4\textwidth]{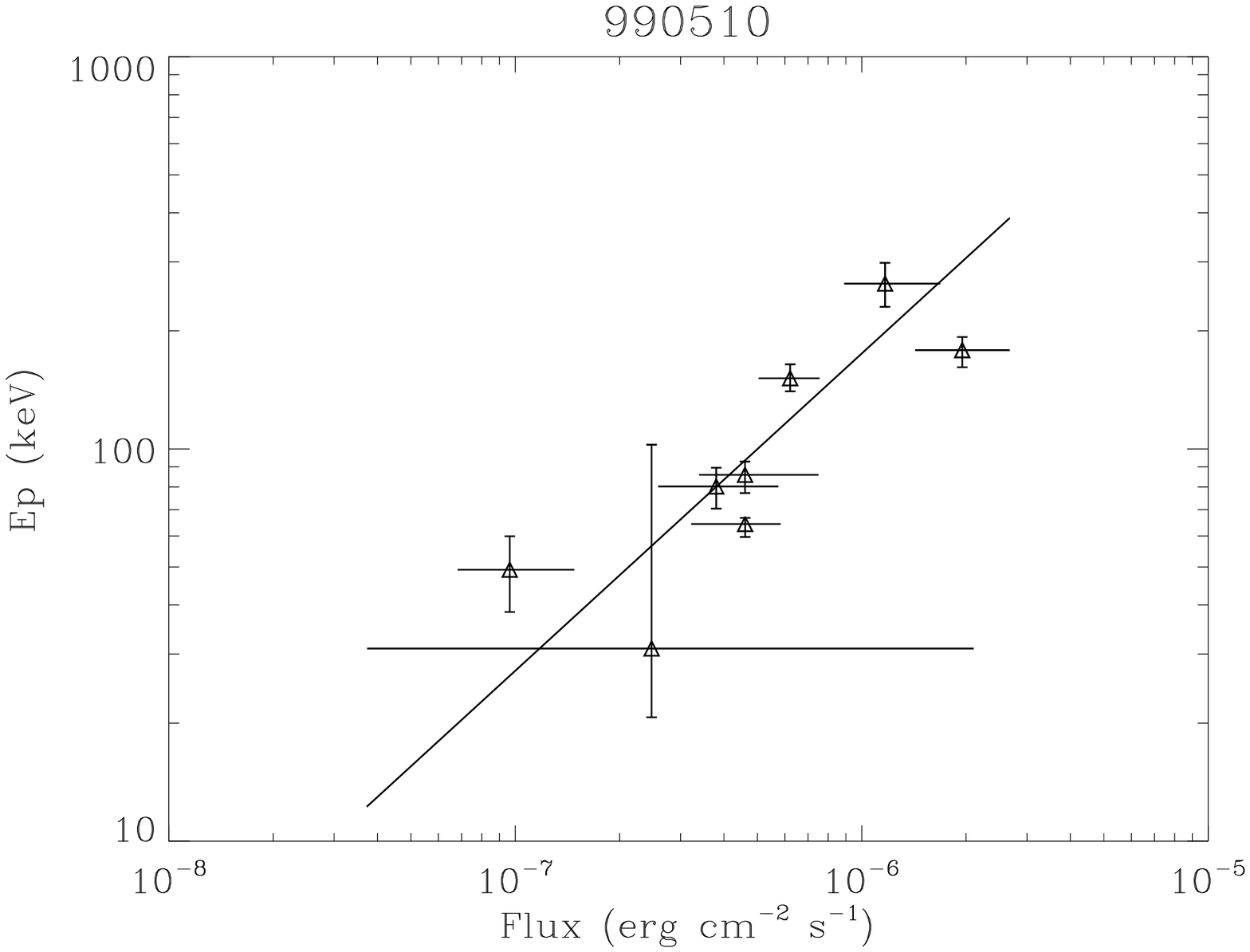}
\end{center}
  \caption{Dependence of the time resolved peak energy, obtained from the best 
fit of the BF to the joint WFC$+$\batse\ spectra, on the 2--2000 keV flux measured 
in the corresponding interval. Also shown is the best fit power-law curve.} 
\label{f:ep-vs-flux}
\end{figure}

\clearpage

%
\begin{figure}
\begin{center}
  \includegraphics[width=.4\textwidth]{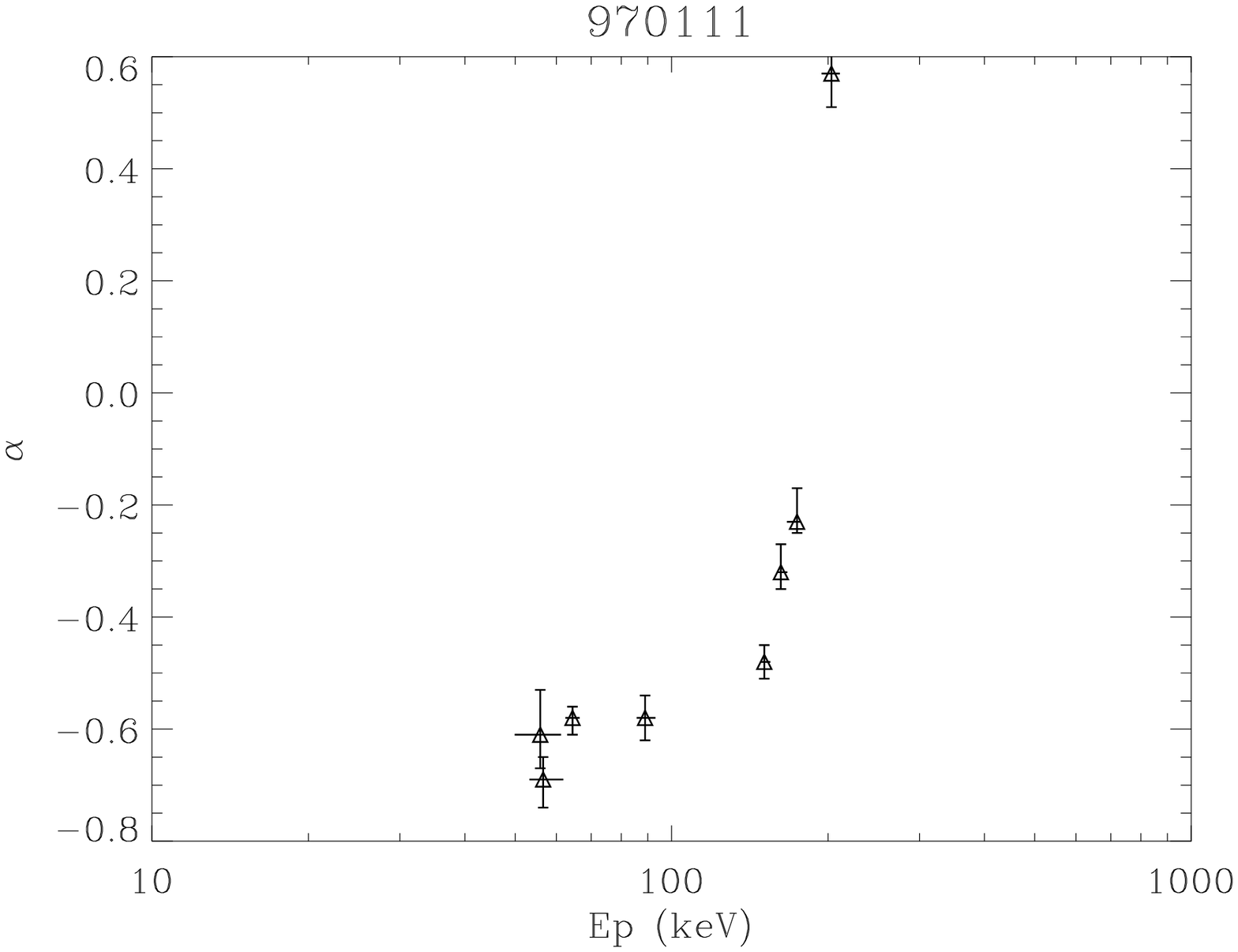}
  \includegraphics[width=.4\textwidth]{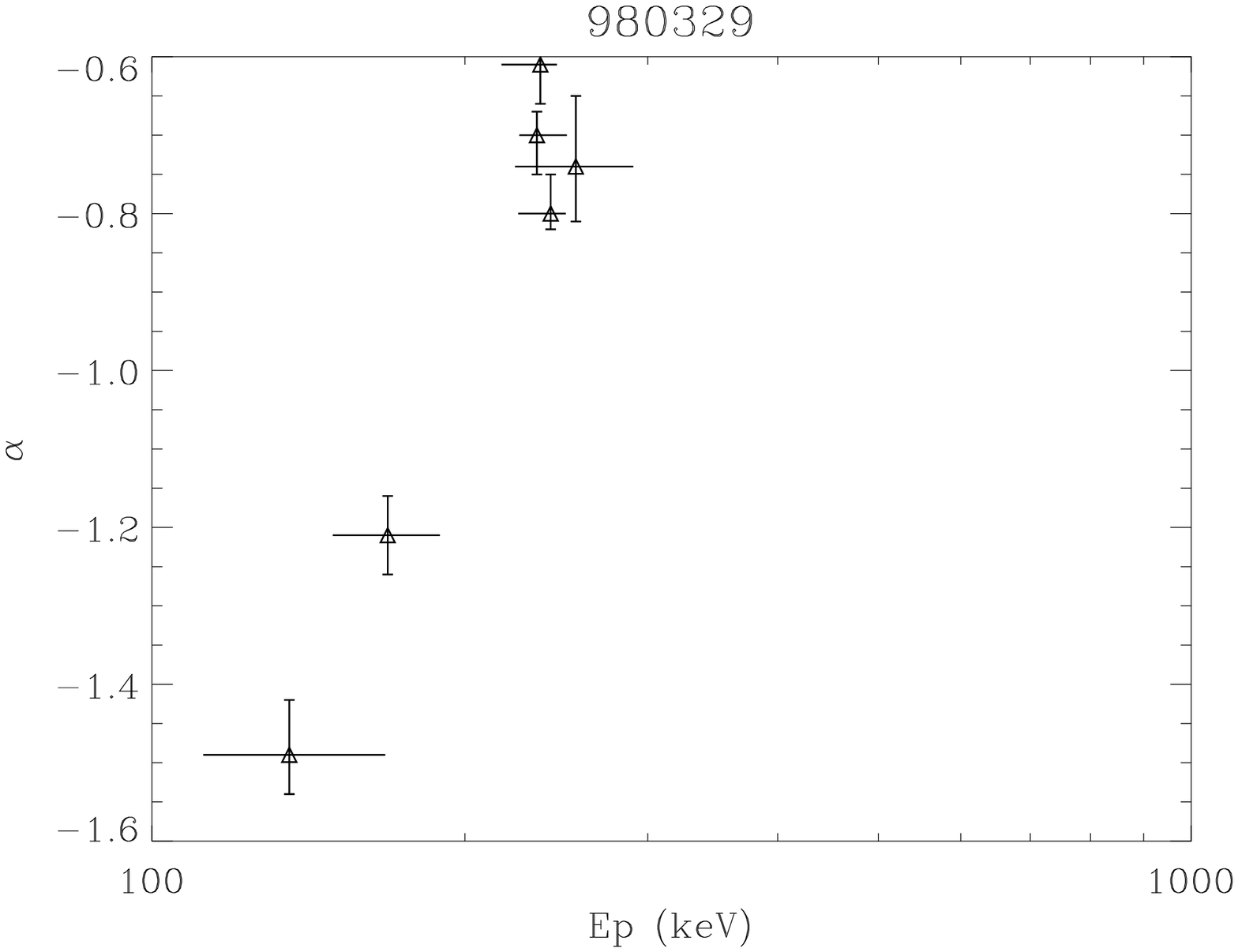}
  \includegraphics[width=.4\textwidth]{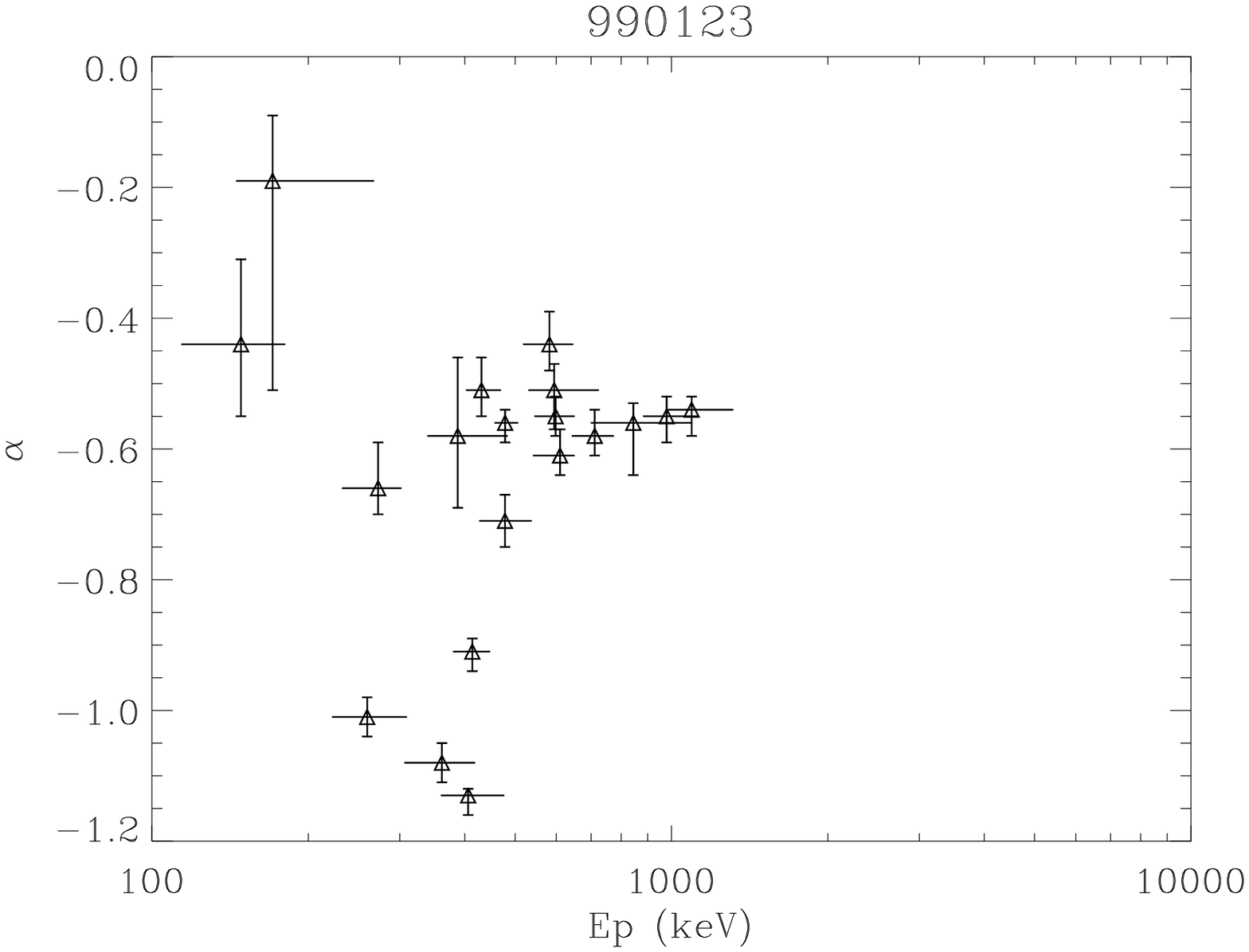}
  \includegraphics[width=.4\textwidth]{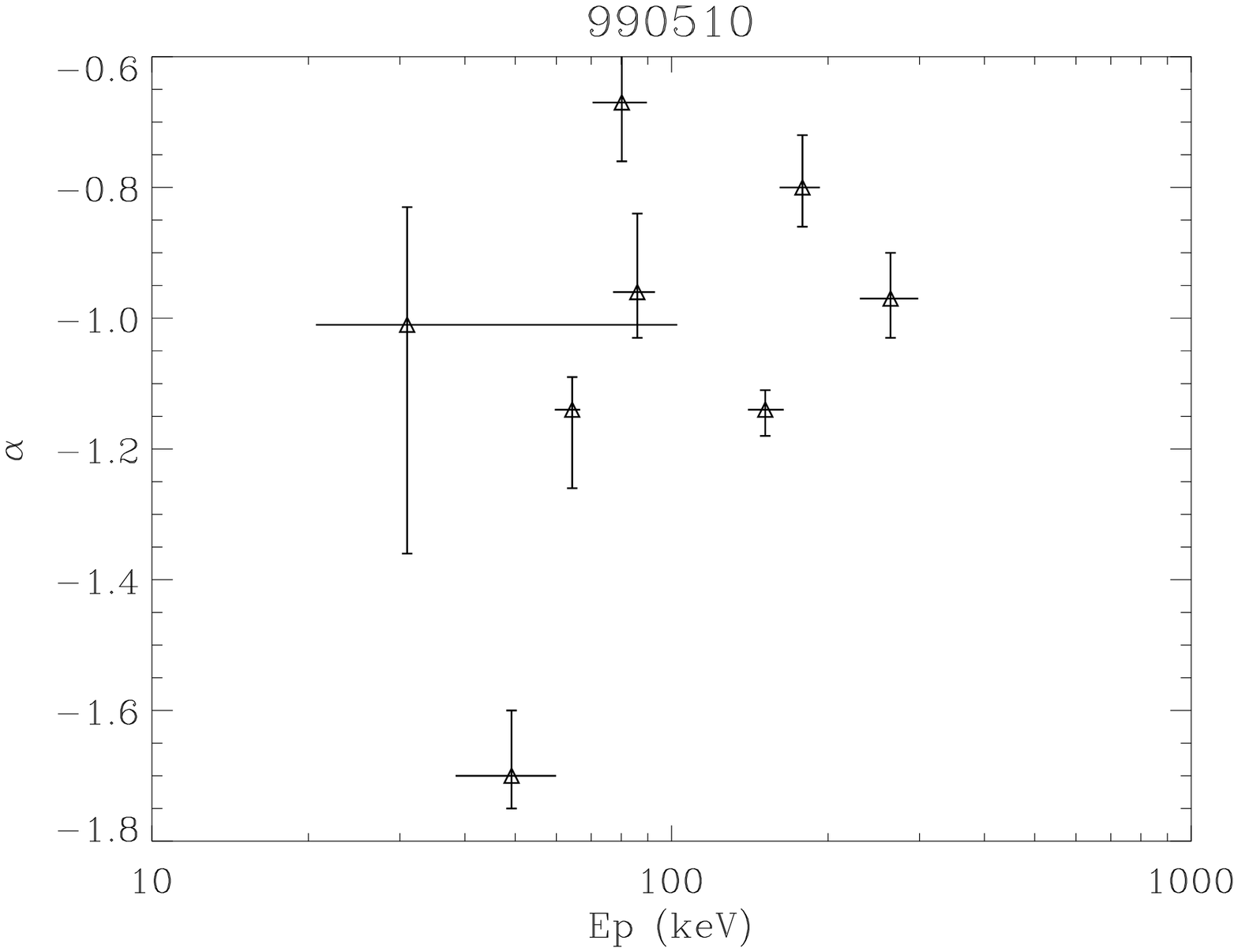}
\end{center}
  \caption{Behavior of the time-resolved low-energy photon index, obtained from the best 
fit of the BF to the joint WFC$+$\batse\ spectra, with the corresponding $E_p$ derived 
in the same spectral interval.} 
\label{f:alpha-vs-ep}
\end{figure}

\clearpage

%
%
\begin{figure}
\begin{center}
  \includegraphics[width=.4\textwidth]{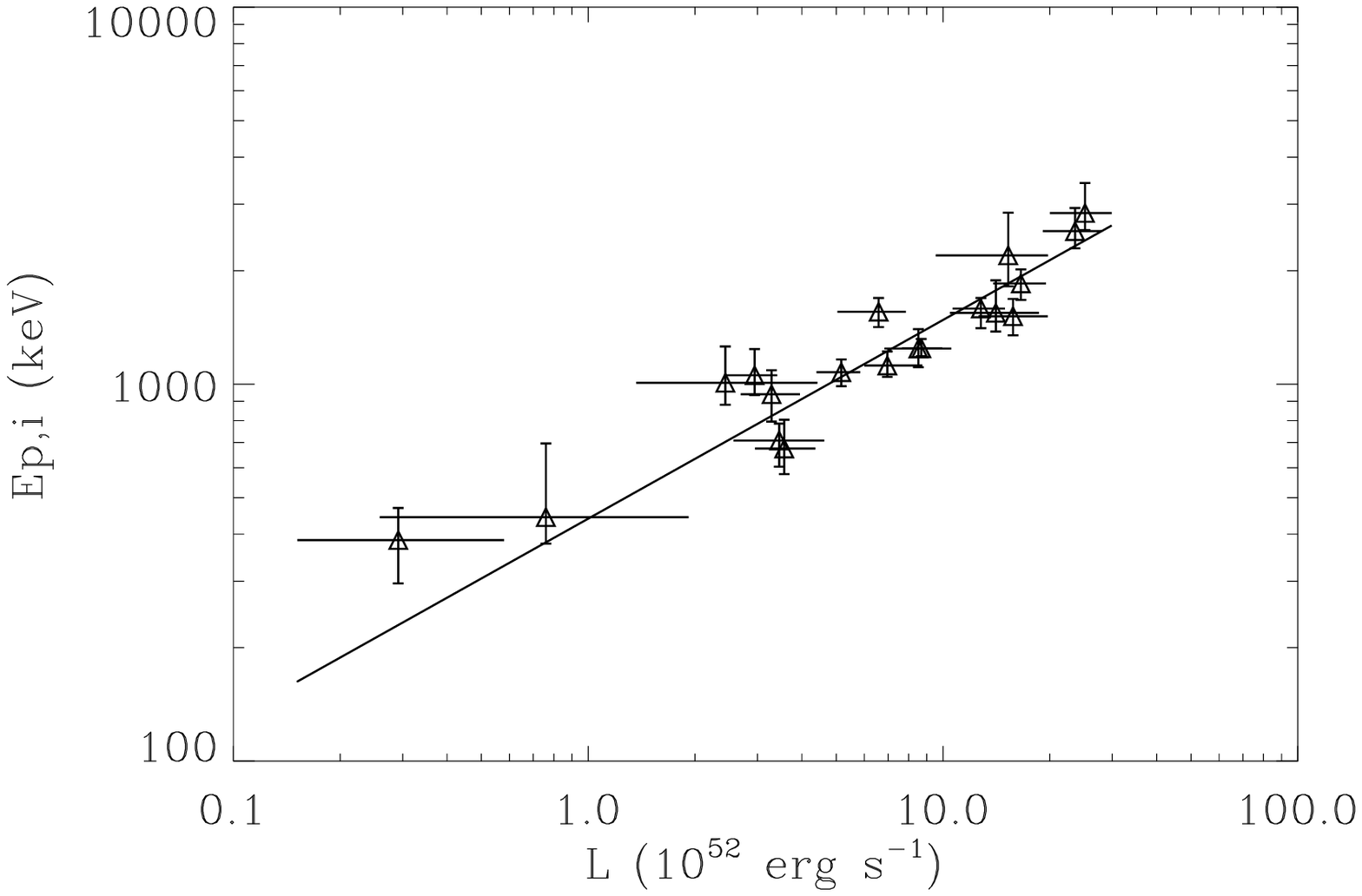}
  \includegraphics[width=.4\textwidth]{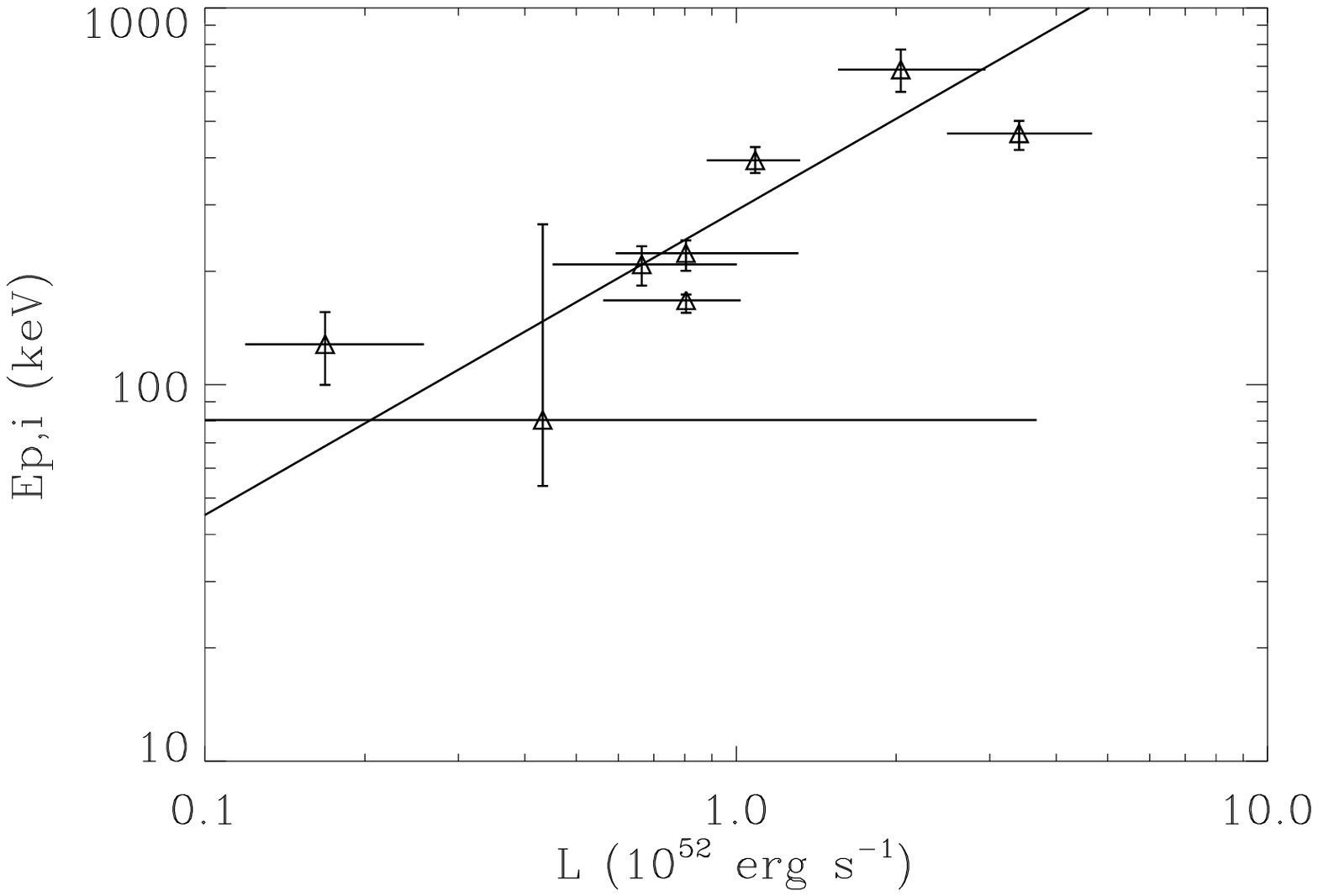}
\end{center}
 \caption{Intrinsic $E_{p,i}$ as a function of the 2--2000 keV luminosity 
for GRBs 990123 and 990510.}
\label{f:Epi-vs-L}
\end{figure}

\clearpage

%
%
\begin{figure}
\begin{center}
	\includegraphics[width=.8\textwidth]{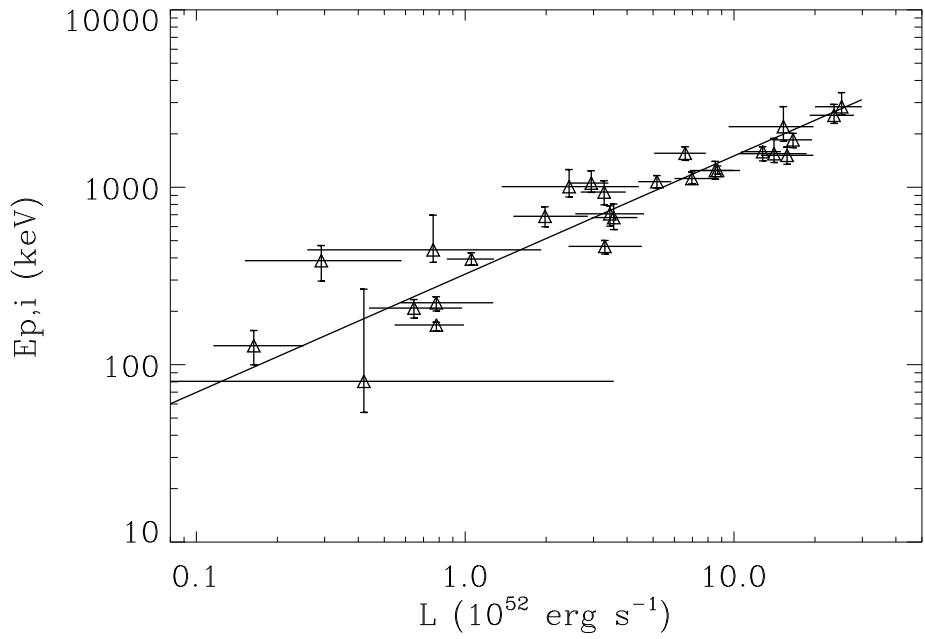}  
\end{center}
 \caption{Intrinsic $E_{p,i}$ as a function of 
the 2--2000 keV isotropic luminosity $L_{\rm iso}$, obtained by merging together all the 
available data for GRB\,990123 and GRB\,990510. 
The continuous line shows the best fit power-law slope (see Table~\ref{t:epi-vs-L}).} 
\label{f:epi-all-vs-L}
\end{figure}

\clearpage

%
%
\begin{figure}
\begin{center}
	\includegraphics[width=.4\textwidth]{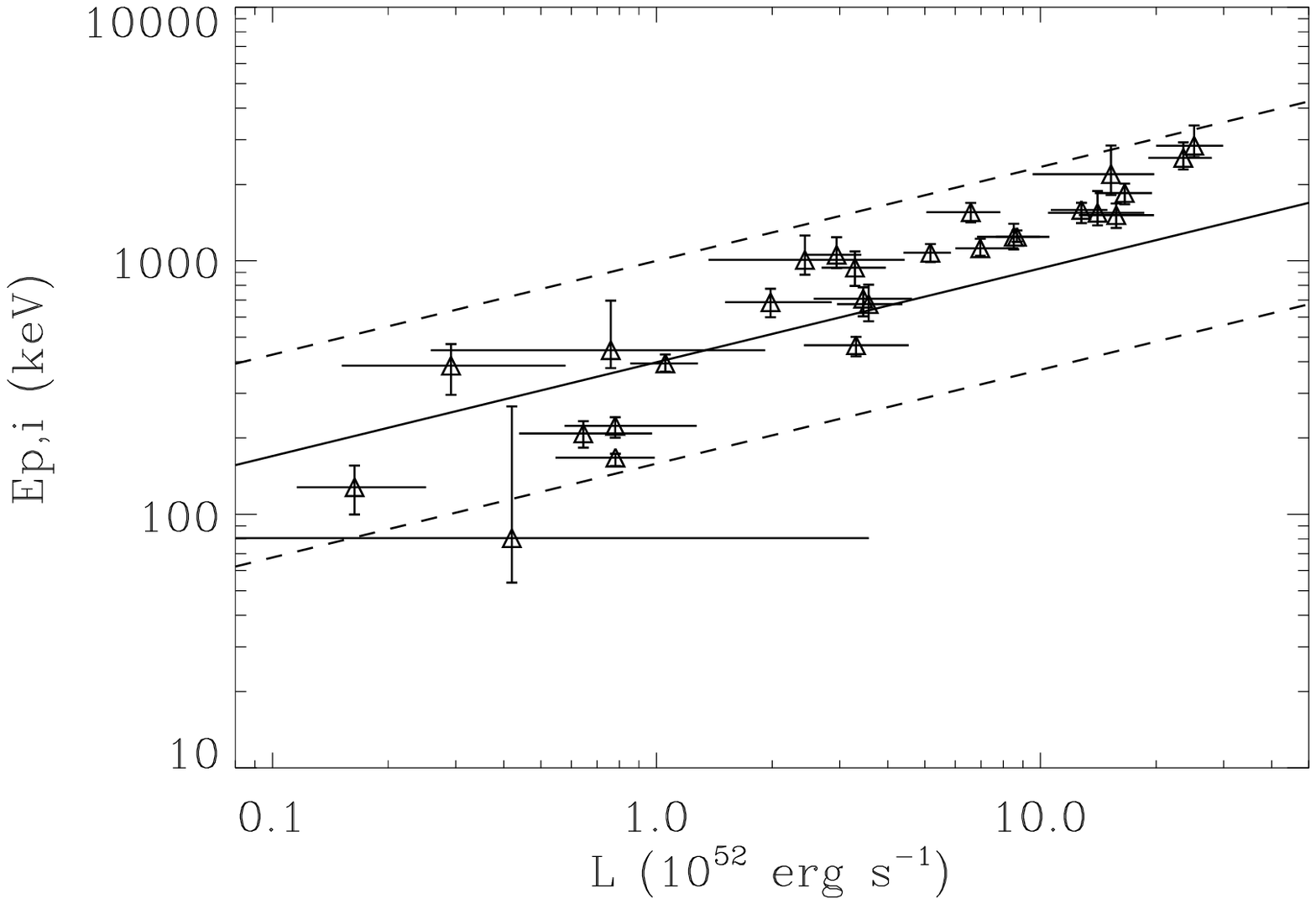}
	\includegraphics[width=.4\textwidth]{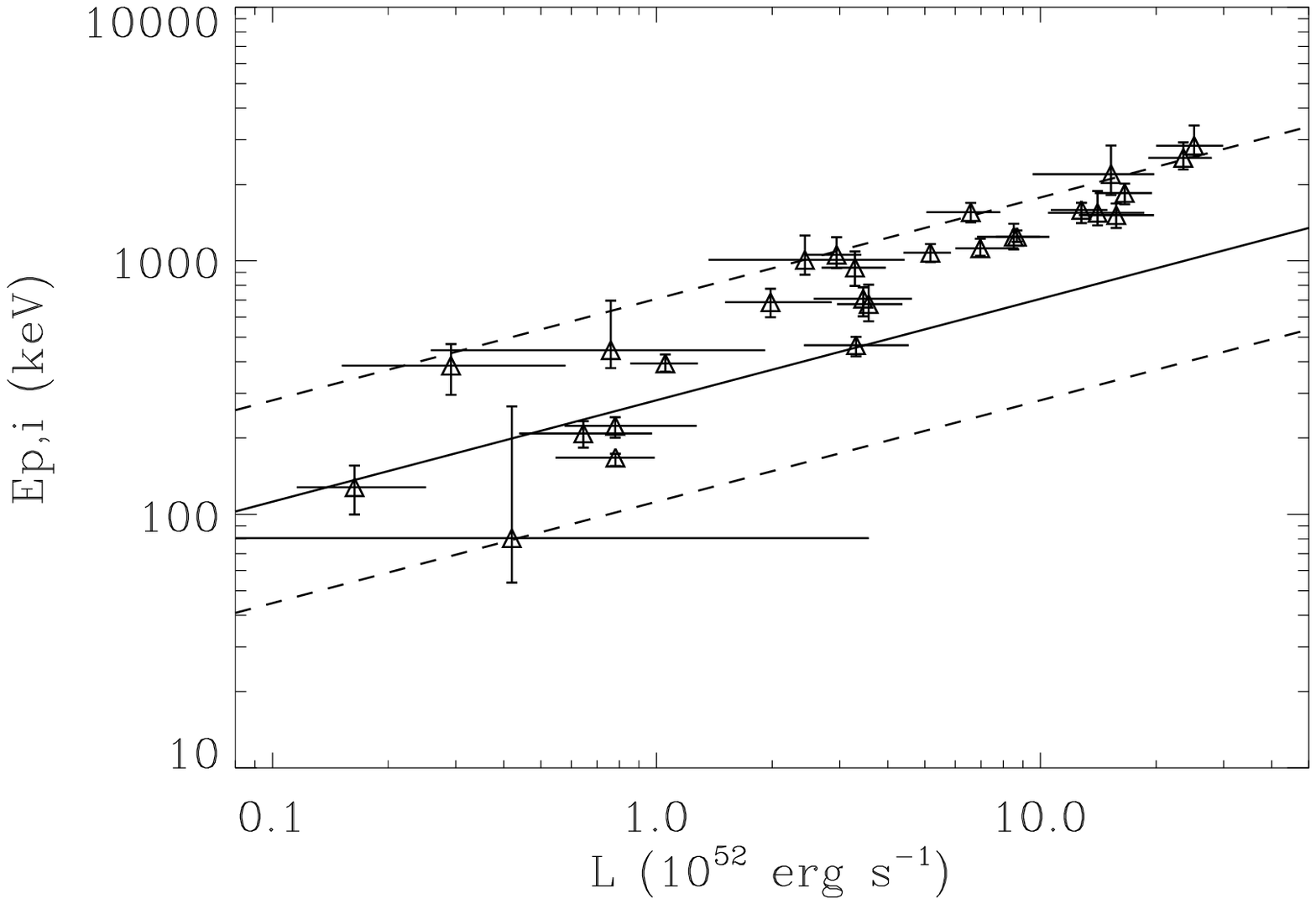}
\end{center}
 \caption{Comparison of the derived $E_{p,i}$ versus $L_{\rm iso}$ with the results 
reported by \citet{Ghirlanda10}.  Left panel:  our results compared with 
the best-fit power-law (continuous line) and the $2 \sigma$ belt (dashed lines) 
obtained using 51 time-resolved spectra extracted from eight GRBs observed with \fermi. 
Right panel: our results compared with the best-fit power-law (continuous line) 
and the $2 \sigma$ belt (dashed lines) obtained by \citet{Ghirlanda10}, using time-averaged spectra of 105 
(pre-\fermi\ plus \fermi) events.}
\label{f:ep-vs-L-comp}
\end{figure}

\end{document}